\documentstyle[11pt,aaspp4]{article}

\def\ea{{\it et al.}~}
\def\etal{{\it et al.~}}

\def\eg{{\it e.g.,~}}
\def\ie{{\it i.e.,~}}

\def \kms{km~$\rm{s}^{-1}$}

\def \cc{$\rm{cm}^{-3}$ }

\newcommand{\be}{\begin{equation}}
\newcommand{\ee}{\end{equation}}

\begin{document}

\title{The Propagation of Magneto-Centrifugally Launched Jets: I}

\author{A. Frank \altaffilmark{1}, Thibaut Lery \altaffilmark{2} 
T. A. Gardiner \altaffilmark{3}, 
T. W. Jones \altaffilmark{4}, D. Ryu \altaffilmark{5}}
\bigskip

\authoraddr{A. Frank\\Dept. of Physics and Astronomy,\\
University of Rochester, Rochester, NY 14627-0171}

\altaffiltext{1}{C. E. K. Mees Observatory, and 
Department of Physics and Astronomy, 
University of Rochester, Rochester, NY 14627-0171; 
afrank@pas.rochester.edu}

\altaffiltext{2}{Department of Physics, Queen's University,
Kingston, ON K7L 3N6, Canada}

\altaffiltext{3}{Dept. of Physics and Astronomy, 
University of Rochester, Rochester, NY 14627-0171; 
gardiner@pas.rochester.edu}

\altaffiltext{4}{Department of Astronomy, 
University of Minnesota, Minneapolis, MN 55455; 
twj@msi.umn.edu}

\altaffiltext{5}{Department of Astronomy and Space Science, 
Chungnam National University, Daejeon 305-764, Korea; 
ryu@sirus.chungnam.ac.kr}

\authoraddr{address used for editorial communication only}


\begin{abstract}
We present simulations of the propagation of magnetized jets. This
work differs from previous studies in that the cross-sectional
distributions of the jets's state variables are derived from analytical models
for magneto-centrifugal launching. The source is a magnetized
rotator whose properties are specfied as boundary conditions. The jets in
these simulations are considerably more complex than the
``top-hat''constant density etc. profiles used in previous work.  We
find that density and magnetic field stratification (with radius) in
the jet leads to new behavior including the separation of an inner jet
core from a low density collar.  We find this {\it jet within a jet}
structure, along with the magnetic stresses, leads to propagation
behaviors not observed in previous simulation studies. Our methodology
allows us to compare MHD jets from different types of sources whose
properties could ultimately be derived from the behavior of the
propagating jets.
\end{abstract}


\section{INTRODUCTION}

Highly collimated supersonic jets are a ubiquitous phenomena 
occurring in many astrophysical environments.  These jets are observed
propagating from sources as diverse as Active Galactic Nuclei (AGN,
\cite{Leahy91}), Young Stellar Objects (YSOs, \cite{Reipurth97}) 
and Planetary Nebulae (PNe, \cite{SokLiv94}).  While considerable
progress has been made in understanding the nature of jets from AGN
and YSOs, there remains considerable debate concerning the nature of
the more recently discovered PNe jets.

The ubiquity of jets in astrophysics has made them a popular subject
for study.  They are excellent laboratories for the study of basic
astrophysical processes (shocks, instabilities, etc.). Their long
dynamical or ``look-back'' times, $t_{dyn} = L_j/V_j$, also make them
ideal astrophysical fossils for studying the evolution of the obscured
and often unobservable central sources, \ie there is the hope in jet
studies that the physics of the central engine can be revealed by
studying the exhaust.  Given the diversity of jet producing
environments there also exists the hope that an underlying unity can be
be found in terms of the fundamental processes which create jets.
Articulating these processes is one of the critical issues facing
astrophysical jet studies.

Accretion disks are believed to play a key role in the physics of both
YSOs and AGN.  In-falling, rotating matter is stored in these disks
until dissipation allows material to spiral inward and feed the
central, gravitating object.  Both YSO and AGN disks are believed to
support strong, well ordered magnetic fields.  The current consensus
holds that these fields are the agents for producing jets in a process
known as {\it Magneto-centrifugal} launching.  In this mechanism,
plasma in the disk is loaded on to co-rotating field lines.  If
conditions in the disk are favorable (\ie field strength and
orientation) the plasma is centrifugally flung outward along the field
lines.  Strong toroidal field components are generated in the flow as
the field is dragged backwards by the plasma inertia leading to
collimation of the wind into into a narrow jet.  We note, however, that
the external medium might also help focus the outflow.
Magneto-centrifugal launching has been studied in detail by many
authors both analytically (\cite{HevNor89}, \cite{Pudritz91},
\cite{Shuea94}, \cite{ostriker}, \cite{Leryea99}) and through numerical
simulations (\cite{OP97a}, \cite{Romanova98}, \cite{Kudohea98}).

In the YSO community two principle flavors of the Magneto-centrifugal
launching model exist.  The first is a pure disk wind model
(\cite{Pudritz91}) in which the jet is generated at the surface of a
Keplerian disk.  The second, called ``X-Winds'' (\cite{Shuea94}),
produces a jet from the boundary layer between the disk and the central
star's magnetosphere.  Other models exist as well (\cite{Goodsonea97})
and there remains considerable debate as too which mechanisms are
obtained in real YSO flows.

While there is an exhaustive literature concerning jet {\it
launching} and {\it collimation}, there has also been considerable
study of jet {\it propagation}.  Propagation studies focus on scales
many orders of magnitude larger (\cite{Reipurth97}) than the region
where collimation occurs.  For example in the work of (\cite{OP97a})
the collimation of the jet was followed out to a height above the disk
of $H = 80 R_i$ where $R_i$ is the inner disk radius.  Since $R_i \le
10 R_*$ ($R_*$ is the stellar radius, \cite{Hartmann98}), the scale of
the simulation was at least 10 times smaller than the smallest scales
on which jets have been resolved and at least $10^3$ times smaller than
the typical scale of observational jet studies.  Much of the
propagation work has been numerical and for both YSOs and AGN much of
it has been have been purely hydrodynamic.  For YSOs only a handful of
MHD studies of jet propagation have been carried out to date
(\cite{Todo92}, \cite{Cerqueira98}, \cite{Frankea98},
\cite{Cenquria99}, \cite{Gardinerea99}, \cite{StoneHardee99}).  If,
however, strong magnetic forces produce the jets then these forces
should effect their propagation downstream.  Unless the fields are
somehow removed, Maxwell stresses should alter at least some
characteristics of the jet's propagation. Recently \cite{Frankea99}
have shown that ambipolar diffusion may be operative in YSO jets in
some part of the flow. However the time-scales involved are such that
changes in jet magnetic fields will only occur for parsec-scale jets.
Flows on time-scales less than $\tau \approx 10^3$ y will not lose
their fields.  In the case of AGN, the ambipolar time-scales are even
larger.  Thus, a proper accounting for the MHD forces in the propagation
of both YSO and AGN jets is needed.  In this paper we focus mainly on
YSOs but our results will be applicable to AGN jets as well.

To date all radiative MHD jet simulations of steady, constant density
``top-hat'' jets have been performed using simple field geometries.
\cite{Cenquria99} showed that jets with purely poloidal 
$\vec{B} = B_z \bf{k}$ topologies did not have propagation
characteristics which differed significantly from pure hydrodynamic
jets. \cite{Gardinerea99} have also found similar results for pulsed
``top-hat'' MHD jets with poloidal fields.  \cite{Frankea98} however,
found that if the field had a strong toroidal ($B_\phi$) component
then the jet head could be strongly effected by the Maxwell stresses 
leading to the production of so-called ``nose-cones''.  
Nose-cones form when post-shock gas 
is restricted from lateral expansion by the axially directed ``hoop
stresses'' associated with strong toroidal fields.  Instead of
back-flowing to form a cocoon, the shocked gas is confined to the head
of the beam in the region downstream of the jet-shock.

The hoop stresses lead to a conical streamlined configuration for the
head i.e. a nose-cone. Such structures were also seen in the early MHD
simulations of AGN jets (\cite{Lindea89}).  In \cite{Frankea98} the
addition of radiative losses, appropriate for YSO jets, caused the
nose-cones to narrow significantly.  In a more extensive set of
calculations \cite{StoneHardee99} found that MHD effects on jet
propagation is strongly dependent on initial field topology.

While these results were promising, there still remains considerable
distance to be traveled in the study of MHD jets.  The principle
issue that must be addressed is that all
the simulations carried out to date is the use of ad-hoc field
topologies.  Unless a force-free configuration is adopted, $J\times{\bf
B} = 0$, Maxwell stresses will act on the jet beam independent of
propagation effects. Thus some effort must be expended in developing
equilibrium configurations for MHD jet simulation initial conditions.
With little to guide them, all modelers have chosen simple topologies
which allow for a simple specification of the required equilibrium.
Frank \ea (1999) used a pure toroidal geometry. \cite{Gardinerea99}
used a pure poloidal geometry. \cite{Cenquria99} used both toroidal and
poloidal as well as force free helical configurations which had to
extend throughout the entire computational domain (jet + ambient
medium).  \cite{StoneHardee99} used helical pressure matched beams in a
variety of configurations.  None of the configurations used in these
papers deviated from the simple constant velocity, constant density
model for the jet beam.  These efforts were necessary for articulating
the basic role of MHD forces in jets, but they do not help establish a
connection between conditions in the jet and the protostellar source (a
protostar and rotating magnetized accretion disk).  What is needed for
use by the broader community is to begin the simulations with jet
cross-sections derived directly from magneto-centrifugal flow models.
That is the goal of the work presented here.

In what follows we present models of MHD jet propagation with initial
configurations in the jet taken directly from the solution of force
balance perpendicular (the Grad-Shafranov equation) and parallel (the
Bernoulli equation) to magnetic surfaces generated by a magnetized
rotator.  Our simulations follow the evolution of jets composed of helical
fields embedded in hypersonic plasmas whose  density and velocity vary with
radius.  Thus our models constitute a further step towards realism in
the theoretical description of magnetized astrophysical jets.  The
goal of this paper is to articulate the basic physics which can occur
in these kinds of jets and to look for differences between the
propagation of jets forming from different kinds of rotators.  We note
that the parameter space of solutions is quite large and in this paper
we present only the first results of this project.  In future papers
we will present a more systematic exploration of parameter space.

The plan of the paper is as follows.  In section 2 we describe the
methods used to construct the initial equilibria and numerically
simulate the flows.  In section 3 we present results of our
simulations focusing on adiabatic, isothermal and radiative cases.  The
next section compares the results with observations. Finally, in
section 5 we present and discuss our conclusions.

\section{Numerical Methods and Initial Equilibria}
\subsection{Basic Equations} 

We numerically integrate the equations of ideal magnetohydrodynamics
(MHD), modified to include the loss of thermal energy due to optically
thin radiative losses.  In cylindrical coordinates these equations take
the following form,
\begin{eqnarray}
\frac{\partial \rho}{\partial t} +
\frac{1}{r} \frac{\partial}{\partial r}(r \rho v_r) +
\frac{\partial}{\partial z}(\rho v_z) & = & 0
\label{bas1} \\
\frac{\partial \rho v_r}{\partial t} +
\frac{1}{r} \frac{\partial}{\partial r}(r \rho v_r^2 - r B_r^2) +
\frac{\partial}{\partial r}(p^*) +
\frac{\partial}{\partial z} (\rho v_r v_z - B_r B_z) & = &
\frac{(\rho v_\phi^2 - B_\phi^2)}{r}
\label{bas2} \\
\frac{\partial \rho v_\phi}{\partial t} +
\frac{1}{r} \frac{\partial}{\partial r}(r \rho v_\phi v_r - r B_\phi B_r) +
\frac{\partial}{\partial z} (\rho v_\phi v_z - B_\phi B_z) & = &
\frac{(B_r B_\phi - \rho v_\phi v_r)}{r}
\label{bas3} \\
\frac{\partial \rho v_z}{\partial t} +
\frac{1}{r} \frac{\partial}{\partial r}(r \rho v_z v_r - r B_z B_r) +
\frac{\partial}{\partial z} (\rho v_z^2 - B_z^2 +p^*) & = & 0
\label{bas4} \\
\frac{\partial B_r}{\partial t} +
\frac{\partial}{\partial z} (v_z B_r - v_r B_z) & = & 0
\label{bas5} \\
\frac{\partial B_\phi}{\partial t} +
\frac{1}{r} \frac{\partial}{\partial r}(r v_r B_\phi - r v_\phi B_r) +
\frac{\partial}{\partial z} (v_z B_\phi - v_\phi B_z) & = & 
\frac{(v_r B_\phi - v_\phi B_r)}{r}
\label{bas6} \\
\frac{\partial B_z}{\partial t} +
\frac{1}{r} \frac{\partial}{\partial r}(r v_r B_z - r v_z B_r)
& = & 0
\label{bas7} \\
\frac{\partial E}{\partial t} +
\frac{1}{r} \frac{\partial}{\partial r}
     \lbrack r (E + p^*) v_r - r B_r (\vec{B} \cdot \vec{v}) \rbrack +
\frac{\partial}{\partial z}
     \lbrack (E + p^*) v_z - B_z (\vec{B} \cdot \vec{v}) \rbrack 
 & = & - \left( \frac{\rho}{\mu} \right)^2 \Lambda(T)
\label{bas8}
\end{eqnarray}
\noindent The total energy and pressure are given by
\begin{eqnarray}
p^* = p + \frac{1}{2} B^2 \\
E = \frac{1}{2} \rho v^2 + \frac{p}{\gamma - 1} + \frac{1}{2} B^2
\end{eqnarray}
\noindent where $\mu$ is the mean molecular weight and  
$B^2 = \vec{B} \cdot \vec{B}$.   Equation
\ref{bas1} and \ref{bas8} represents conservation of mass and energy
respectively.  Equations \ref{bas2} - \ref{bas4} represent
conservation of momentum.  Equations \ref{bas5} - \ref{bas7} represent
the induction equation.  The energy conservation equation includes a
source term, $n^2 \Lambda (T)$, (where $n = \rho/\mu$ is the number
density), which models radiative losses in the optically thin limit.
We use the Dalgarno-McCray ``coronal'' cooling curve
(\cite{Dalgarno1972}).  A ``floor'' temperature of $T_f = 10^4$ K is
set such that gas can not cool to lower values.  The fluid is assumed
to be an ideal gas, where $\gamma$ is the ratio of specific heats.
Other relevant quantities are the sound speed $c = \sqrt{\gamma
p/\rho}$, the Alfv\'en speed parallel to the magnetic field $v_a =
\sqrt{B^2/\rho}$ and the plasma {\it beta} parameter, $\beta = 2
p/B^2$.

In addition to the hyperbolic equations represented above an additional
constraint is imposed via the condition of flux conservation,
\begin{equation}
\frac{1}{r} \frac{\partial}{\partial r} B_r +
\frac{\partial}{\partial z} B_z = 0.
\label{divb}
\end{equation}
Using these equations we model the propagation of a magnetized jet
through a constant density, constant pressure magnetized ambient
medium.  The initial conditions for the jet, \ie its cross sectional
distribution of $\rho, ~p, ~\vec{v}$ and $\vec{B}$, are calculated via
the Given Geometry Method of (\cite{Leryea98}, \cite{Leryea99},
\cite{LeryFrank99}).  We describe this method and the equilibrium MHD
jet solutions it produces in the next section.

\subsubsection{The Model}
The jets we inject into the computational grid are taken directly from
a (simplified) model of the magneto-centrifugal launching/collimation
process.  The model, known as the Given Geometry Method (GGM:
\cite{Leryea98}, \cite{Leryea99}, \cite{LeryFrank99}) allows asymptotic
MHD jet equilibria to be linked directly to the properties of a
rotating source. The GGM assumes a time-independent, axisymmetric
flow.  It further simplifies the problem of magneto-centrifugal
launching/collimation by assuming that the nested magnetic flux
surfaces defining the flow (labeled by the variable $a$) possess a
shape which is known a priori inside the fast critical surface.  The
fast surface defines the locus of points beyond which the flow is 
kinetic energy dominated. The flux surfaces are assumed to be conical
and, as an additional simplification, an equilibrium across the
surfaces is assumed at the Alfv\'en point which yields an equation
referred to as the Alfv\'en regularity condition. This condition is 
not a criticality condition since the Alfv\`en point is not strictly
a critical point.

The flow properties must be determined by solving for the equilibrium
of forces parallel and perpendicular to the magnetic surfaces (the
former described by using the Bernoulli equation for a polytropic
equation of state and the latter is solved via the Grad-Shafronov
equation). The equilibrium parallel to the surfaces takes the form of
criticality conditions at the two other (fast and slow) MHD critical
points.  This corresponds to differential form of the Bernoulli
equation on constant $a$ with respect to $\rho$ and $r$ vanishing at
the critical points.

In the general case the GGM yields five integrals of motion that are
preserved on any axisymmetric magnetic surface $a$.  Two of the
integrals are given as boundary conditions in the model.  These are the
angular velocity $\Omega(a)$ and an entropy (or ploytropic) factor $Q(a)$.  
These are supplied as a model for the source rotator.   We note that
the entropy paramter Q(a) can be decribed as follows: The density
$\rho$ is related to the pressure $p$ by a polytropic equation of
state, $p =  Q(a) \rho^{\gamma}$ where $\gamma$ is the polytropic index
and Q the polytropic constant that is related to the entropy. This
assumption replaces consideration of energy balance and is meant to
simply represent more complex heating and cooling processes (See, for
example, \cite{VT} for more general equations
of state).  Changing Q(a) changes the local thermal energy balance in
the flow.

The Alfv\'en regularity condition together with the criticality
conditions then determine the three other unknown integrals:  namely
the specific energy $E(a)$; the specific angular momentum $L(a)$;
the mass to magnetic flux ratio $\alpha(a)$.  Far from the source
(large z) the flow becomes cylindrically collimated.  In this
asymptotic regime the jet is assumed to be in pressure equilibrium
with an external medium. The pressure matching condition along with
with the Grad-Shafronov and Bernoulli equations are all solved in
the {\it asymptotic cylindrically collimated regime}.

We note that in the GGM the source (e.g. the accretion disk) is not
explicitly described since it is point-like.  Instead the shape of the
magnetic field lines defined by the flux function a(r,z) is specified
out to the fast magnetosonic point, but not its angular distribution.
The rotation rate $\Omega(a)$ and the polytropic parameter $Q(a)$ are
then specified on the field lines. The shape of the magnetic field
lines from the fast magnetosonic point to the fully collimated region
is not specified. The strong toroidal component which develops in the
wind and the fully collimated jet develops mainly due to differential
rotation and the interia of mass on the field lines.  This is similar
to other disk wind models \cite{OP97a}.

\subsubsection{Numerical Solutions}

Inside the fast critical surface, the variables calculated in the
numerical procedure are the energy $E$ and the radii and densities at
the three critical surfaces, ($r_s$, $r_f$, $r_A$, $\rho_s$, $\rho_f$,
$\rho_A$).  In the asymptotic cylindrically collimated 
regime,
the jet is entirely defined by this set of $r$ and $\rho$ and all other
physical quantities can be derived from them. For the numerical
calculations, the equations have been reformulated as ODEs or converted
from algebraic conditions into ODEs as functions of the flux surfaces
$a$.  The system consists of eight differential equations and the
numerical solutions are obtained by initiating the integration of the
system from the axis. Given the input parameters $Q(a)$, $\Omega(a)$,
$\alpha_0$, $\gamma$, and $\rho_0$, all the critical positions and
densities can be numerically obtained using analytical formulae (see
\cite{Leryea98}).  We further constrain the solution to be
super-Alfv\'enic and super-fast-magnetosonic on the axis in the
asymptotic region.  

\subsubsection{Classes of Jet Equilibria}

\placefigure{fig:OMEGA}
In our approach the most important aspect of the source rotator is is angular
rotation profile.  We focus on this aspect of the source because it most clearly
connects to different scenarios of magneto-centrifugal launching/collimation. 
Profiles of angular velocity of the source rotator considered in this
paper are shown in Fig.~\ref{fig:OMEGA}. The pure {\em Keplerian}
rotator (dashed line) starts with a constant rotation close to
the axis, as in the rigid body case (dot-dashed lines), but then
follows a {\em Keplerian} profile. The {\em Multi-component} (solid
line) case also starts with a rigid rotation corresponding, for
example, to an axial ordinary wind. The angular velocity then doubles
its value in order to model a jet rotating more rapidly than the star
in an intermediate region between the ordinary wind and the Keplerian
disc wind that follows. Note that the angular velocity is always
sub-{\em Keplerian} in the intermediate region.  For all the rotation
laws, the axial value of the angular velocity $\Omega_0$ is set to
unity in the Figure, and the radius is normalized to the size of the
jet. For reference we also show the profile of a solid body rotator
though we will not consider this model in the simulations.
Fig.~\ref{fig:ASYMP} also shows that a
return poloidal electric current flows back inside the jet for both the
{\em Keplerian} and the {\em Multi-component} jets.  

As shown in \cite{LeryFrank99}, it is possible to derive an
approximate analytical solutions of the model in the cylindrical region.
It has been found that the density can be expressed as a function of $r$ 
and of the first integrals as $\rho(r) \approx {C \alpha(r)}/{\Omega(r) r^2}$
where $C$ is a constant. The asymptotic poloidal velocity of the flow
can also be derived and is given by $v_z(r) \approx \Omega(r) r / C$.
Therefore the velocity increases with the angular velocity while the 
density decreases. This explains why the density drops when $\Omega$
is important in the inner part of the jet for the Multi-component case,
while it increases afterwards in the Keplerian rotation regime.
The velocity roughly follows an opposite behavior with respect to the
angular velocity. More detailed analysis of these equilibria are 
given by \cite{LeryFrank99}.

\subsubsection{Identification of Basic Features}

\placefigure{fig:ASYMP}
The quantities that define the jet in the cylindrically collimated
regime, are plotted in Fig.~\ref{fig:ASYMP} for pure {\em Keplerian},
{\em Multi-component} and constant rotations.  The {\em Keplerian} and
{\em Multi-component} models will be used as input for the numerical
simulations.  The $z$ and $\phi$ components of velocity and magnetic
field are represented together with the density $\rho$ and the net
electric current $I_C$, as functions of the relative radius
(normalized to the jet radius). The length scale is the jet radius,
the density is normalized to its value on the jet axis $\rho_0$, and
the non-dimensional velocities refer to the fast magnetosonic velocity
$v_f^2 = c_s^2 + v_A^2$ on the axis, $c_s$ being the sound speed.  The
magnetic field is normalized to $\sqrt{\rho_0}\ v_f$.

The most important features in these graphs are the variations of the
toroidal component of the magnetic field and of the density.  Note that
the region near $R = .3R_j$ is dominated by the magnetic pinching
force, or {\em hoop stress}, where $B_\phi$ is maximum. The gas
pressure is important at radii less than this value in order to
maintain the equilibrium and is the origin of the large density
gradients in this region.  We denote the high density region centered
on the axis as the {\em core}, and the lower density outer regions as
the {\em collar}.  Note that the bulk of the jet's momentum resides in
the core.  Hence we expect this portion of the beam to penetrate more
easily into the ambient medium during the jet's propagation while the
collar will be more strongly decelerated.   Fig.~\ref{fig:ASYMP} also shows that a
return poloidal electric current flows back inside the jet for both the
{\em Keplerian} and the {\em Multi-component} jets.  More detailed
analysis of these equilibria are given by \cite{LeryFrank99}.

Thus Keplerian and Multi-component jets are characterized by a
{\em dense, current-carrying core, carrying most of the momentum}, 
and are surrounded by a collar carrying an internal return current.

\subsubsection{Scaling for the Simulations}

The input parameters of the model can be selected so as to
qualitatively reproduce observed situations. Given the properties of
the jet-emitting object, \ie its radius $R_{\ast}$, its temperature
$T_{\ast}$, the total mass loss rate $\dot M_{\ast}$, the base density
$n_{\ast}$, the magnetic field $B_{\ast}$, the factor $Q_{\ast}$ and
$\gamma$, it is possible to deduce the dimensionless parameters
$\overline \Omega$, $\overline Q$, $\overline \alpha_0$.  The parameter
$\overline \alpha_0$ can be a-posteriori related to the mass loss rate
$\dot M_{\ast}$, $R_{\ast}$, and the magnetic field $B_{\ast}$. So we
define $ Q_{\ast}\equiv {2 k T_{\ast} n_{\ast}}/{(m_p
n_{\ast})^{\gamma}}$, $\alpha_{\ast} \equiv {\dot M_{\ast}}/{4 \pi
R_{\ast}^2 B_{\ast}}$, and $\Omega_{\ast}\equiv
\sqrt{{GM_{\ast}}/{R_{\ast}^3}}$.  All those quantities are
non-dimensionalized to reference values by setting $\overline Q \equiv
Q_{\ast}/Q_{ref}$, $\overline \alpha_0 \equiv
\alpha_{\ast}/\alpha_{ref}$ and $\overline \Omega \equiv
\Omega_{\ast}/\Omega_{ref}$.  The entropy $\overline Q(a)$ is assumed
to be constant across the jet.  In the present paper, we have chosen to
model YSO jets with different rotation laws using typical values for
TTauri stars as presented by \cite{bertout}. At the base of flow, we
deduce the corresponding dimensionless input parameters:  ${\overline
Q} =0.87$, ${\overline \Omega} =2$, ${\overline \alpha_0} =0.7$, and
$\rho_0=5.10^{-7}$.  Major quantities of reference are then given (in
CGS) by $R_{ref} = 10^{15}~cm$, $n_{ref}=250~cm^{-3}$, $v_{ref} =
10^{7}~cm~s^{-1}$ for Young Stellar Objects. $V_ref$ is simply a 
canonical speed for YSO jets which we use to set the
scales in the simulations.

\subsubsection{Comparisons with other models} 

A detailed comparison of the present model with previously published
studies has been given by \cite{LeryFrank99}.  Here we report only the
most important conclusions.  As with Ferreira (\cite{ferr3}), it is
possible to show that the GGM yields a minimum mass loss rate injected
in the jet which has a lower limit and can not be arbitrarily small.  These
results also agree with Ostriker (\cite{ostriker}) and Lery
\etal(\cite{LHF}) who conclude that the optical jet may represent only
the densest part of the total outflow.  We obtain a fast magnetosonic
Mach number, (which also corresponds to the Alfv\'enic Mach number on
the axis), between 2 and 4. This range corresponds to what
(\cite{cam182}) has found for his model for low-mass protostellar
object. The corresponding jets have low fast magnetosonic Mach-numbers
$M_A\simeq 2$. By taking into account an accretion disc around the
stellar magnetosphere, Fendt \& Camenzind (\cite{fendt}) also find a
fast magnetosonic Mach-number to be $2.5$. This results does not,
however, appear to be a general statement about MHD jets since there
exist models with larger values (Sauty \etal(\cite{sautytsing}) and
Trussoni \etal(\cite{trussoni})).  Finally, the analytical results
given by Shu \etal(\cite{shu95}) agree with those of the GGM model in
terms of jet structure. We note however that despite the similarity of
the analytical results,  neither the {\em Multi-component} (or the 
{\em Keplerian}) case can be seen as equivilent to the X-wind model.
Note in particular that our model does not describe the physical
processes occurring at the source itself, \ie at the surface of the
disk or the disk-star boundary.  

\subsection{Numerical Method and Implementation}

A detailed description of the numerical code can be found
in references given below.  Here we simply state the code's
most salient features.  Specifically, the method we use to solve
equations \ref{bas1}-\ref{bas8} is explicit, finite element (volume),
up-winded, conservative, $2^{\mathrm{nd}}$ order accurate, and
total variation diminishing (TVD).  TVD stands for Total Variation
Diminishing.  This refers to the ability of the code to caputure strong
discontinuities in the flow without producing spurious oscillations.
TVD methods are part of a general class of "High-Resolution"
codes which solve the hydro or MHD equations in conservative form
by using a Gudinov method (ie solving the Riemann problem at every
grid interface) and including sophisticated algorithms for limiting
the fluxes through cell boundaries to keep the solution monotone.
More detail concerning High Resolution methods can be found in
\cite{Leveque98}.  The code is conservative up to machine accuracy,
ensuring that it will accurately capture shock strengths and speeds.
It has been well tested in standard 1-D shock tube tests as well as
multi-dimensional stability calculations. Various manifestations
of the code have been reported in the literature including its
1-dimensional (1-D) cartesian form (\cite{Ryu1995-1D}), its 2-D
cartesian form (\cite{Ryu1995-2Dxy}), and its 2-D axisymmetric
(cylindrical coordinates) form (\cite{Ryu1995-2Dcyl}).  The TVD
property is ensured in the same way as was done originally by Harten
for the Euler equations in (\cite{Harten1983}).  In the 2-D versions
of the code, multidimensionality is handled through the use of Strang
splitting (\cite{Strang}).  The cooling is applied in a first order
fashion.  Finally, the crucial and problematic issue of maintaining
$\vec{\nabla}\cdot\vec{B}=0$ is accomplished with a staggered grid
approach (\cite{DivBmeth}).

Each simulation was carried out in cylindrical coordinates
$(r,\phi,z)$ with axisymmetry and inversion symmetry across the $z=0$
plane.  We follow a quarter meridional plane 
($r\ge0$, $z\ge 0$, $\phi=0$) with
$512 \times 2048$ grid cells.  The jet radius spans $64$ grid cells.
Thus our simulations follow the jet propagation for $Z = 32 R_j$.
In Table 1 we present a set of important physical parameters
for the simulations presented below.  While we will use dimentionless
variables in most of the description that follows Table 1 allows
the reader to compare the physical scales in the simulations
with observations. 

In what follows we express all distances in terms of $R_j$ and all
times in terms of the magnetosonic crossing time $\tau = R_j/M_{f}$
where $M_{f} = \sqrt{v_a^2 + c^2}$ (note: for our initial conditions
the magnetosonic speed and the fast mode speed are identical). We
utilize outflow boundary conditions at the outermost radial and axial
boundaries.  In cylindrical coordinates the $r = 0$ line is
necessarily a reflecting boundary.  During the tests which we have run
we have found little evidence of incorrect reflections from the $r=0$
line due to the coordinate singularity, though it must be admitted
that this problem plagues all numerical codes in cylindrical
coordinates with axisymmetry.  Inversion symmetry through the $z=0$
plane dictates the use of reflecting boundary conditions at the $z=0$
plane. In some simulations we have found that waves propagating inward
from the outermost radial boundary (caused by the bow-shock
propagating off the grid) led to a slow compression of field at the
base of the jet at late times.  We wished to avoid the use of
logarithmic grids hence we  suppressed 
the inflow of material at the outer radial boundaries problem by
injecting a slow  ($M_s \sim 1.2$) wide angle flow at at the base of the
grid.  This advected the material off the grid and kept the field from
being overly compressed near the jet inlet at the base of the grid and
the imposed flow had no effect of the propagation of the jet far
downstream.

In each simulation we inject the jet into the computational domain via
2 layers of  ``ghost-zones'' below the base $z = 0$ of the grid.  
The jet properties are read into
the grid from data files provided by the Given Geometry Model
described above.  The density and pressure in the ambient medium are
copied from values in the last radial zone of the jet: $\rho_a =
\rho_j(R_j), ~ p_a = p_j(R_j)$.  In addition the ambient medium is
given a pure poloidal magnetic field ($\vec{B}_a = B_{a,z}\hat{k}$)
whose magnitude is also taken from the last radial zone of the jet
$B_{a,z} = B_{j,z}(R_j)$.  Since $B_{j,\phi}(R_j) = 0$ our jets are in
radial pressure balance the with the ambient medium.

In order to articulate the basic dynamics inherent to the flows we
have run three classes of model for both the {\em Keplerian} and {\em
Multi-component} jet.  In our Adiabatic models we have set $\gamma =
5/3$ and turned off the cooling source term.  In our Isothermal models
we have set $\gamma = 1.001$ and turned off the cooling term.  In our
Radiative models $\gamma = 5/3$ and the cooling source term was turned
on.  We have run both radiative and isothermal models as consistency
checks as well as to allow us to model jets with different Mach
numbers.  For reasons explained above the equilibria provided by the
Given Geometry model yielded jets of low magnetosonic number ($3 < M_f
< 5$).  We wish to model jets with lower 
temperatures. This could be accomplished
by scaling down both the pressure and magnetic field such that the force
balance was maintained.  In this way we were able to model jets with Mach
numbers of order ($6 < M_f < 9$).

The cross sectional variation of $\vec{B}_j(r)$ in the jet presents a
problem in terms of initial conditions.  This is a general difficulty
which all attempts to model the evolution of magnetized jets must
confront.  Flux conservation, $\vec{\nabla} \cdot \vec{B} = 0$,
demands that any discontinuities in the field must be associated with
current sheets (which are the cause of field kinks at MHD shocks).
Attempts to initialize simulations of magnetized jets propagating into
magnetized ambient media must deal with the likely mismatch of field
topologies and magnitudes at the head and sides of the jet when the
simulation is first switched on.  The use of cylindrical coordinates
eliminates the problem for the $B_\phi$ component.  While an initial
discontinuity in the toroidal component may produce transients, it
will not violate flux conservation.  Thus we must only deal with the
$r$- and $z$-components of the field.  In our simulations we solved the problem
by continuing the 
$z$-component of the jet field into the ambient
medium.  Thus for $r < R_j$, $B_{a,z}(r) = B_{j,z}(r)$. Since $B_z$ is
relatively weak, 
$B_z^2/2 \ll p$, the gradient in the ambient field produces little mass motion.

In order to test the effect of the initial conditions on the observed
behavior (\ie transients) we have run a series of models which began
with the jet and ambient conditions joined smoothly via a hyperbolic
tangent function.  The smoothing length $h$ was varied from $h = .5 ~
R_j$ to $h = 9 ~ R_j$.  We found that the long term behavior of the
jet was unaffected by the choice of h.  We also note that this version
of the code produces a relatively strong boundary layer at the
jet/ambient gas interface at distances far from (well behind) the head
of the jet.  While such layers are to be expected due to unresolved
instabilities (mainly Kelvin-Helmholtz modes) we found the effect was
partially attributable to the treatment of transverse wave modes
in the code. We performed a number
of tests to ensure that changes in the flow variables in the boundary
layer were not affecting the results.

%
\section{Results}

In the this section we present the results of the simulations. We
provide a description of the behavior seen in the models along with
attempts to understand the underlying physics.

\subsection{Basic features} 

Several features are common to almost all of the simulations.  In the
input equilibria the core-collar structure is always present with
gradients of jet variables between the jet core and collar.  When the
equilibrium jet encounters the external medium, the various elements of
the equilibrium are shocked.  This creates two bow-shocks and a cocoon.
The bow-shock closest to the axis takes the form of a nose cone.
Intrinsic instabilities develop in the inner part of the shocked core,
as well as in the cocoon. The annotated Figure ~\ref{fig:cartoon} sums
up this section by showing the set of common features on a
characteristic {\em Multi-component} jet.  We now focus on the
propagation characteristics of the two types of input equilibria.

\subsection{Keplerian Rotator} 

To breifly review {\em Keplerian} rotators produce jets with a nearly
constant velocity cross section.  The mass density is stratified with
a high density {\it core} surrounded by a lower density {\it collar}.
The core-collar density ratio for the present jet is relatively low:
$\rho_{core}/\rho_{col} = 1.67$, (this is also the density ratio
between the core and the ambient medium, $\eta = \rho_j/\rho_a$).  The
toroidal magnetic field in the jet reaches its maximum value just at
the outer edge of the core.  As we shall see, the coupling of higher
density in the core with the strong magnetic stresses along the
core/collar boundary dominates the propagation characteristics of the
entire jet.

\subsubsection{Keplerian Rotator: Adiabatic Jet}

\paragraph{Propagation} 
In Fig.~\ref{fig:DenAdKep} we present gray-scale maps of the density
evolution of an adiabatic jet driven by a {\em Keplerian} rotator.  In
the first frame, taken at $t = 5.3 \tau$, the classic
jet-shock/bow-shock pair are apparent. The bow-shock accelerates the
ambient gas while the jet-shock decelerates the jet material.  The
speed of the jet head or bow-shock is $v_h \approx 77$
\kms.  This speed is relatively constant throughout the 
simulations. \cite{Frankea98} derived a formula for the bow-shock 
speed which accounted for magnetic pressure in the beam.  Using the 
familiar result for hydrodynamic jets, $v_{ho} =
\frac{v_j}{1~+~1/\sqrt{\eta\nu}}$, ($\nu$ is the ratio of 
bow-shock and jet head 
radii $R_h/R_j)$ the MHD bow-shock speed is. 
\begin{equation} 
v_{h} = v_{ho} {
{1~-~\sqrt{\frac{1}{\eta\nu}~-~\frac{p^*}{\rho_j v^2_j}
(1~-~\frac{1}{\eta\nu})}}\over{1~-~\frac{1}{\sqrt{\eta\nu}}}}, 
\label{BigE}
\end{equation}
\noindent If we take $\nu = 1$ then 
this equation gives $v_h \approx 70$ ~\kms with magnetic
pressure accounting for approximately 4\% of the momentum flux driving
the shock.  As was noted in \cite{Frankea98}, the higher velocity of
the jet head seen the simulations can be attributed to the aerodynamic
effect of streamlining the jet head via MHD hoop stresses 
(an $\nu$ \emph{effect}). 
The nose-cone shape which develops reduces the drag on the jet head
increasing its velocity relative to a more blunt jet head which would
occur in a pure hydrodynamic simulation.

In the first frame of Fig.~\ref{fig:DenAdKep} we already see the
effect of the core/collar structure on the jet propagation.  The
higher density core, confined by the magnetic hoop stresses, maintains
its structural integrity on the downstream side of the  
jet-shock. It is noteworthy at this early time that the core appears to propagate
ahead of the rest of the beam.  This is to be expected purely from
momentum considerations as equation \ref{BigE} predicts a $\Delta v =
10$ \kms~difference in the speed of the core and collar bow-shocks.
Detailed examination of the simulations also shows that at these early
times there is little material flowing from the jet into the cocoon.
This is apparent in Fig.~\ref{fig:KAvhead} which shows the poloidal
plane velocity vectors for the head of the jet. The origin of this
effect lies, once again, in the relative strength of the toroidal
field in the core and collar.  What material does flow into the
cocoon comes primarily from the outer most regions of the collar ($r >
.75 R_j$).  Note that there is no transverse motion in the shocked
core material.  We attribute this to magnetic forces.  In the region
where material is flowing in the radial direction the magnetic field
has an average value that is $1/3$ that at the core/collar interface.
A better means of judging the relative strength of the field comes
from examination of the plasma parameter $\beta$.  Downstream of the
jet-shock the core collar interface has $\beta = 3$ while the collar/ambient
interface has $\beta= 23$.  Thus fields can exert stronger stresses
along the core restricting its lateral expansion.

The remaining frames of Fig.~\ref{fig:DenAdKep} show the distance
between the jet and bow-shock 
continues to grow. Note that as the jet evolves the core never loses
its identity. It acts, essentially, as a {\it jet within a jet}.  At
later times we see secondary shocks developing within the shocked core
as well as vortex shedding at the head of the jet at radii consistent
with the core/collar interface. In particular, by the second frame we
see what appears to be a second jet-shock 
forming inside the core as it pushes through the ambient gas.  Note
also that at later times the shocked core material takes on the
familiar nose-cone morphology seen in top-hat MHD jets with strong
toroidal geometries.  These features emphasize the apparent
independence of the core's propagation characteristics relative to the
rest of the jet.  Thus our results show that the core/collar dichotomy
appears to control the main features of the jet beam via gradients in
inertia and Maxwell stresses.

\paragraph{The Lateral Expansion}
The second notable feature in the adiabatic {\em Keplerian} jet
simulations is the large scale mass expulsion event which occurs in
the third frame in Fig.~\ref{fig:DenAdKep}.  The plasma driven
laterally (in the $r$ direction) in this event is composed entirely of
shocked collar material.  At early times the cocoon is fed solely
through the outer annuli of the jet as Fig.~\ref{fig:KAvhead}
demonstrated.  The hoop stresses in the material in the collar at
smaller radii  are, however, too strong to allow plasma to stream transversely 
into the cocoon.  To
see this explicitly consider the radial momentum equation where terms
involving $v_\phi$, $B_r$ are ignored and we also ignore variations in
$z$.
%
\begin{equation}
\frac{\partial \rho v_r}{\partial t} +
\frac{1}{r} \frac{\partial}{\partial r}(r \rho v_r^2) =
- \frac{\partial}{\partial r}(p) 
- \frac{\partial}{\partial r}\left(\frac{1}{2} B_z^2+B_\phi^2\right)
- \frac{(B_\phi^2)}{r}
\label{hoopstres}
\end{equation}
%
The last term on the right is the hoop stress.  The second to last
term is the magnetic pressure.  From Fig.~\ref{fig:ASYMP} it is clear
that for $r > .3 R_j$ both the gas and magnetic pressure gradients are
negative. Thus in these regions the hoop stress opposes the pressure
forces and acts to constrain lateral expansion of the flow.

At later times however two features occur which alter the balance of
forces.  First as more material builds up immediately behind jet
shock, both the gas pressure and magnetic pressure increase relative
to the magnetic tension.  Second, and most importantly, the jet-shock
becomes distorted, tipping towards direction of jet propagation.  The
jet-shock becomes conical with the vertex of the cone pointing into
the undisturbed beam.  Thus the jet-shock
becomes oblique relative to the
un-shocked mass flux.  The shock conditions for velocity lead to
the following expression for the post-shock radial velocity
\begin{equation}
v_{r} \approx \frac{1}{4} v_j \cos(\theta)\sin(\theta)
\label{vzps}
\end{equation}
where $\theta$ is the angle between the jet-shock 
and the $z$ axis and $v_j$ is the jet velocity relative to the jet-shock.
For $\theta < 90^o$ the post-shock gas acquires a
significant $v_r$ component. Material in the beam is refracted away
from the axis.  The momentum flux in the outward radial ram pressure,
$\rho v_r^2$, is able to overwhelm the magnetic tension force leading
to a large scale expulsion of material into the cocoon.

\paragraph{The Shocks} It is difficult to isolate the processes which
cause the bending of the jet-shock into a conical shape.  The dynamics
at the jet head are highly non-linear and time-dependent and it is not
obvious if the change in shock geometry is an amplification of events
downstream where the flow pattern is quite complicated or if the
distortion can be 
linked to events upstream.  Close inspection of the
simulations gives the impression that the distortion of the jet-shock
occurs after a pinch wave reflects off the axis just upstream of the
jet-shock in the un-shocked beam.  The origin of the pinch appears to
come from the slow expulsion of material into the cocoon {\it prior}
to the third frame.  In the early evolution the cocoon distorts the
flow of ambient gas {\it behind the bow-shock}, \ie the cocoon
represents an obstacle which the post-bow-shock flow must stream
around.  As the shocked ambient material streams over the cocoon it
becomes transonic.  Its return to parallel streaming along the jet
boundary can only occur via an additional shock.  This feature is
apparent at $z \approx 8 R_j $ in frame 2 of Fig.~\ref{fig:DenAdKep}.
Such flow patterns are well known to areodynamicists as they are common
in aerofoil theory (\cite{Ramm90}).  It appears that the pinch wave is
generated just downstream of this additional shock and may be
attributable to the higher pressures generated on the surface of the
jet.

After the large mass expulsion event the jet-shock appears to relax to
a configuration where it is perpendicular to the z-axis  ($\theta =
90^o$) and the flow into the cocoon is reduced.  As the expelled
material curls back towards the jet beam, however, it impinges on the
jet surface and another, strong pinch is generated.  At the
end of the simulation (after the jet head has moved off the grid) we
find the jet-shock becoming distorted yet again leading, perhaps, to a
second mass-shedding event.

\paragraph{The Magnetic Effects} In Fig.~\ref{fig:FieldAKep} a we show
the magnetic field structure in the adiabatic {\it Keplerian} jet.  The
Figure shows the poloidal ($\vec{B}_p = B_r \hat{e}_r + B_z \hat{e}_z$)
magnetic field lines and the toroidal field ($B_\phi$).  The field
clearly traces out the main features of the flow described above: the
nose-cone at the jet head; the mass ejected behind the jet shock; the
pinch wave occurring where the ejected mass is swept back onto the jet
beam.  It is noteworthy that it is the toroidal field which articulates
these structures most clearly.  This is appropriate as the toroidal
field dominates in the jet providing much of the force which shapes the
jet dynamics.  Note that the Figure indicates that the magnetic pitch
$B_\phi/|B_p|$ increases behind shocks. This is a general feature of
helical fields in jets.  In a fast MHD shock only the component of the
field parallel to the shock face is strengthened via compression (the
parallel component will scale as $\rho$).  Thus in a jet with a helical
magnetic topology shock waves act to comb out the field leading to
enhanced toroidal fields (and hoop stresses) in the post-shock regions
(\cite{GardFrank99}).  We note however that shear in the flow
will also lead to significant strengthening of the field via stretching 
of field lines.

Our results for the {\em Keplerian} model show that the behavior of a
jet with a more realistic 
initial density, pressure and magnetic field
structure leads to propagation characteristics which have not been seen
in previous hydrodynamic jet simulations.  As we shall see this theme
is repeated in all the simulations.

\subsubsection{Keplerian Rotator: Isothermal/Radiative Jet}

\paragraph{The Propagation}
In Fig.~\ref{fig:DenRdKep} we present gray-scale maps for the density
evolution of the radiative {\it Keplerian} rotator jet.  Recall that
the radiative model begins with modified initial conditions compared
with the adiabatic or isothermal Keplerian simulation.  In order to
keep the initial temperature at $T = T_o\approx 10^4 ~K$ we scaled
down all the radial distributions of all variables in the jet except
$v_z$.  This had the additional effect of producing a jet with a
higher Mach number, $M_f \approx 7$. We note that the basic features
seen in the radiative simulation are quite similar to those seen in
the isothermal model ($\gamma = 1.001$).  Thus for brevity we do not
present the isothermal results.  The only notable difference between
the two models is the width of the bow shock. This can be understood
purely in terms of the opening angle $\theta_c$ of the Mach cone for a
supersonic flow, $\theta_c = \sin^{-1} (1/M_f)$. We see a wider
opening angle for the lower Mach number isothermal flow as expected
from the relation for $\theta_c$.

The dynamics of both the isothermal and radiative simulations are
dominated by the loss of pressure support between the jet- and
bow-shocks.  In the isothermal model this occurs because $P \propto
T_o$ where $T_o$ is a constant equal to the ambient temperature.  In
the radiative model the gas behind both the bow- and jet-shocks are
driven to temperatures of $T = (3/16)(\mu/k) V_{s}^2 \approx 10^5 K$
where ($s$) refers to the shock speeds.
The cooling time for a jet with $n_j = 100$ \cc is $t_c = .25
T/(n_j\Lambda (T)) \approx .2 (\tau) = 5$ years.  Thus we expect the
post shock gas at the head of the jet to cool effectively and for the
dynamics to be, essentially isothermal.  This is confirmed by
consideration of the first frame in Fig.~\ref{fig:DenRdKep} which shows
by $t = 1.38$ the two shocks have already collapsed on to each other
producing a thin shell.  The densities and magnetic field strengths in
the shell are high with $1000$ \cc $< n < 6000$ \cc and $100 ~\mu G ~<
B < 300 ~\mu G$.  Unlike purely hydrodynamic radiative shocks, MHD
radiative shocks possess a theoretical limit for the post-shock
compression.  This occurs because of magnetic pressure exerted by the
component of the field perpendicular to the shock normal. Equating ram
pressure and magnetic pressure allows a simple form of the maximum
post-shock density to be derived
(\cite{HollMcKee79}),
\begin{equation}
n_m = \frac{(2 m_h)^{1/2} (n_j)^{3/2} v_s}{B_{\phi0}}.
\label{maxcomp}
\end{equation}
Note the above use our scaling for the field. 
For the {\em Keplerian} jet $n_m \approx 1000$ \cc which is in good
agreement with the simple prediction above.  The higher densities
achieved in the simulation come from the pinch force induced
compression on the axis

\paragraph{The Shocks}
The effect of the radial stratification in both the density and
magnetic field (the core/collar structure) are already apparent at the
first frame of the simulation just as in the adiabatic models.  In the
radiative simulation however, the bow-shock quickly assumes a 
pointed, cusp-like shape.  This is due to the higher density in the
jet core and the magnetic pinch forces from the strong toroidal
field. Compared with the adiabatic simulations described above the
head of the jet assumes what might described as a ``bullet'' shape
rather than a nose-cone.  In \cite{Frankea98} significant streamlining
was observed in the radiative simulations compared with adiabatic
ones.  This was attributed to the loss of thermal energy and, hence,
the increased effectiveness of magnetic stresses.  Here we see a
similar effect which is enhanced by the increased ram pressure in the
jet core relative to the collar.

We also see a mass shedding event in these simulations though it is far
weaker than what occurs in the adiabatic models.  Frame 2 of 
Fig.~\ref{fig:DenRdKep} shows the initialization of the event.
As in the adaibatic models a secondary bow shock wave generated
by his event leads to pinching of the jet beam and a downstream
distortion of the jet shock.

Note the cusp which appears in the {\it jet-shock } in the first frame
of the simulation.  While all features on the axis of an axisymmetric
simulation must be taken with some suspicion, a close examination of
the simulation data reveal a straight-forward explanation for this
structure.  The strongest post-shock field values in the jet head occur
just downstream of the jet-shock at a radius where the pre-shock field
is a maximum.  Recall that this occurs just on the outside edge of the
jet core $r \approx .3 R_j$.  This is also where $\beta$ drops to its
lowest value, $\beta \approx
.5$.  Thus magnetic stresses dominate the plasma at this location in
the jet head.  At radial positions just inward of the point where
$\beta = \beta_{min}$ we find $v_r$ obtaining is maximum inward
(negative) value.  This radially inward flow is apparent in
Fig.~\ref{fig:KCvhead} in which we present the velocity vectors at the
head of the jet.  Thus, at positions immediately downstream of the
jet-shock magnetic forces squeeze and compress the jet core.  Given the
strong cooling, the relation $P\propto\rho$ is approximately valid and
the axial location of the pinch is a local pressure maximum.  The
downstream pinch can communicate upstream with the jet-shock face thus
producing the bulge or cusp which faces into the oncoming material in
the beam.  Note that this feature was not seen in the adiabatic models
because there the post-shock gas pressure was high enough to inhibit
the strong pinch.  In fact, $\beta$ will always increase across an
adiabatic shock.  If we write the post-shock compression as
$X=\rho_2/\rho_1$ with the subscripts $1$ and $2$ corresponding to 
pre-and post-shock conditions respectively then (\cite{Priest86}),
\begin{equation}
\frac {\beta_2}{\beta_1} = \gamma M_f^2 \left(1 - \frac{1}{X^3}\right) - 
\frac{1}{\beta_1}
\left(1 - \frac{1}{X^2}\right). 
\label{betrat}
\end{equation}
%
The equation above shows that for strong adiabatic shocks
($M_f \gg 1$, $X \approx 4$)
$\beta_2 > \beta_1$.  It is only when the post-shock thermal energy is
lost to radiation that the post-shock magnetic forces can dominate.

The magnetic field structure in the jet is shown in the bottom panel of
Fig.~\ref{fig:FieldRKep}.  The most prominent feature of the field
configuration is the compact size and high field strengths in the jet
head.  The effect of the pinch wave is clearly apparent upstream of the
jet head.  The field also shows the effect of the weaker mass shedding event
which occurs in this model.  Note 
the isolated loops of $B_\phi$ and strong distortion of $B_{pol}$
behind the cusp in the bow-shock. It is also notable that the turbulence and
multiple instabilities which are seen in most radiative jet simulations
do not occur here.  As \cite{GardFrank99} have found for their pure
poloidal simulations this is one of the principle effects of strong
magnetic fields.  Thus if YSO jets do contain strong embedded fields
then one must consider what their effect on the morphology of the HH
objects should be.

\subsection{Multi-component Rotator} 

The structural differences between the {\em Keplerian} and {\em
Multi-component} jet originate primarily in the differences in density
and velocity cross sections.  As we saw in Fig.~\ref{fig:ASYMP} the
{\em Multi-component} jet has three structural elements: a high
density core; a low density inner collar; a moderate density outer
collar.  The ratio of the peak density in the core to minimum density
in the inner collar is $\rho_{max}/\rho_{min} = 100$.  This is almost
two orders of magnitude higher from what 
is obtained in the {\em Keplerian} jet.  In addition, the velocity in the jet
peaks in the inner core just at the point where the density drops with
$V_{max}/V_{min} = 1.7$.  In the last section we discussed how the
(milder) cross sectional variations in the {\em Keplerian} jet had
important dynamical consequences for its propagation characteristics.
Thus we expect the more extreme radial variations in the {\em
Multi-component} jet will likely effect the dynamics in more extreme
ways.

\subsubsection{Multi-component Rotator: Adiabatic Jet} 

\paragraph{Propagation and ``Peel-off''} Fig.~\ref{fig:DenAdMC}
shows the evolution of the {\em Multi-component} jet through four
gray-scale maps of $log$ density taken at different times in the
evolution of the simulation.  The most prominent feature in the flow is
what we have termed the ``peel-off'' of the jets' outer collar.  As the
jet propagates down the grid, the outer collar develops a strong
radial velocity component.  As the outer collar expands sideways it
is decelerated and develops into a large scale vortex.  This 
is somewhat similar to what was seen in the mass expulsion event seen in
the {\em Keplerian} case.  The high density core of the jet continues
its forward propagation driving through the ambient medium at high
speed and quickly pulls away from the decelerated collar.  By the end
of the simulation the ``naked'' core has 
propagated 
far downstream where it encounters 
the ambient medium in a manner unaffected by the
outer collar.

The origin of the peel-off appears to reside in the 
density stratification of
the jet.  First we note that since the peel-off occurs early in the
simulation we must be suspicious of it as an artifact of the way the
simulations are initiated.  Given the nature of this study it is
difficult to circumvent the need to begin our simulations with a fully
formed jet as this is the point of the project.  As was noted in
section 2 experiments in which the smoothing length between the jet and
ambient conditions was varied revealed no change in the propagation
characteristics.  Even when the initial head of the jet was smoothly
joined with ambient medium over a length of many jet radii we found the
development of the peel-off was only delayed. The outer layers always
developed their transverse motion and the evolution was identical to
models with shorter or no smoothing transition.  Thus, while the
development of this feature may be a transient, it is a highly robust
one.  Consideration of the dynamics inherent to stratified jets such as
these however allows one to infer the mechanism driving the peel-off.
We focus on the {\it jet} shock.  The highest post-shock pressures
occur behind the highest velocity regions of the jet.  This occurs in
the low density inner collar.  Since $\eta(r) < 1$ in this region the
jet-shock is relatively strong and is pushed back into the jet deeper
than in either the core or outer collar.  The oblique geometry of the
inner shock generates a strong transverse flow in both positive and
negative radial directions.  This can be seen in Fig.~\ref{fig:MAvhead}
which shows the poloidal flow vectors. In addition the variation of the
the inner shock produces a finger of high pressure gas which reaches
back into the jet in the low density inner collar.  A strong radial
pressure gradient is established which drives the outer collar away
from the core much like splitting wood with an axe.  Once the sideways
expansion begins, the ram pressure of the ambient medium (in the frame
of the jet) continues to divert the flow of the outer collar.

The difference in jet propagation speeds between the {\em
Multi-component} and {\em Keplerian} jets is also dramatic. The
velocity of the bow-shock at the end of the simulation is $V \approx
100$ \kms{}  which is a $25\%$ increase over the propagation speed of the
{\em Keplerian} jet.  This difference can be attributed two effects.
First, the {\em Multi-component} jet has a higher value of $\eta$ in
the core relative to the ambient medium ($\eta \approx 6.8$ for the
{\em Multi-component} jet).  From equation \ref{BigE} this translates
into a relative propagation velocity difference of $14\%$.  The excess
in propagation speeds above this is likely 
to be attributable to a second
effect - the streamlining of the jet head.  Once the outer collar peels
away from the core, the jet presents a smaller and more streamlined
head to the ambient medium allowing it to propagate at higher speeds.
This can be seen by comparing the bow-shock opening angles for the {\em
Multi-component} and {\em Keplerian} jets.  Note the streamlining of
the head of the naked core also comes via the strong magnetic pinch
forces at its outer radial edge and at late times the core also
develops the familiar nose-cone morphology.

\paragraph{The Instabilities} The development of strong instabilities
in the core of the {\em Multi-component} jet is another notable
characteristic of the simulations.  Once the core is exposed we see
periodic pinches in the beam.  As the instabilities evolve they expand
radially and develop a arc-like shape.  At later times these arcs
become swept backwards by shear in the beam. Further evolution leads to
a loss of their sharp edges and individual arcs begin to merge.  From
detailed consideration of animations of the simulations it appears that
the instabilities first appear near the head of the naked core.  Only
at later times as the peel-off the outer layers continues do they
appear immediately downstream of the peel-off region.  This is most
likely an indication of where the perturbations driving the
instabilities occur.

We have performed a stability analysis of the {\em Multi-component}
magnetic configuration (Lery \cite{lery3},\cite{LeryFrank99}).
A global normal mode stability analysis was performed using the
same method as in Appl, Lery \& Baty (\cite{alb}). In these two papers the stability of
magnetized astrophysical jets with respect to modes driven by the
electric current density distribution was addressed. The results
show that the current driven (CD) instabilities grow rapidly on time
scales of order of the Alfv\'en crossing time in the jet frame and
that they are likely to modify the magnetic structure of the jet.
Since they are internal modes (see Appl, Lery \& Baty (\cite{alb}))
the CD instabilities should not disrupt the jet.  In the present work,
we have focused on the pinch mode because of the axisymmetric nature
of our calculations.  In 3-D it is likely that the kink mode may
play a role as well but should not break the integrity of the jet.
This is because the jet is super-fast and should see its boundary as a
rigid wall.  Consequently, the instabilities should be mainly internal,
the jet would not be disrupted. The instabilities should certainly
be expected to change the jets magnetic configuration drastically .

The analysis shows that the strong pinch, (or sausage), mode is
mainly due to large gradients of the density and magnetic field.
In Fig.~\ref{fig:stab}, we have plotted the dispersion relation for
different values of the external pressure surrounding the jet.  We have
adopted the standard temporal approach where the axial wavenumber is
real and the imaginary part of the complex frequency corresponds to
growth rate.  Wave-numbers are given in units of inverse jet radius
and growth rate is normalized to the inverse Alfv\'{e}n time.  It has
been found that the location of the peak mainly depends on the magnetic
distribution in the jet. The short-$k$ cut-off is due to the finite
size of the jet radius that has an external boundary that behaves as
a rigid wall for Mach numbers larger than unity (see Appl, Lery \&
Baty (\cite{alb})).  We find that pure magnetic instabilities driven
by electric current develop on rapid Alfv\'{e}n time-scales.  Also,
Fig.~\ref{fig:stab} clearly shows that when the external pressure
increases the jet becomes more unstable. This is precisely what we
observe when the core becomes naked downstream of the peel-off region.
It also explains why the instabilities do not develop as rapidly for
the {\em Keplerian} case where the density and pressure gradients
are less important.

From the simulations it appears that at later times waves driven off
the peel-off region seed the instabilities while at earlier times the
seeds occur via shocks at the jet head.  As can be seen in
Fig.~\ref{fig:DenAdMC} the pinching instabilities on the axis have a
wavelength of approximately $\lambda \approx.5 R_j$.  Note however the
presence of a second characteristic wavelength which runs along the
surface of the core.  This feature, which appears as an envelope
encompassing the shorter wavelength modes, has $\lambda\approx 3 R_j$.

The stability analysis of {\em Multi-component} jets
shows that the unstable modes that should grow the most rapidly have a
wavelength of about 3 jet radii for the collar and half the jet radius
for the core.  These results have been reported by
\cite{LeryFrank99}.  They are in good agreement with the simulations,
and can be also compared to observations. For example, the jet of HH34
presents a mean knot separation of $3.4~r_{jet}$ as given by Burke
\etal ({\cite{burke}).  Thus, the present results suggest that these
instabilities could be at the origin of the knotty structure of a large
number of jets as seen, for example, in HL Tau, HH1, HH30 and HH34 (Ray
\etal\cite{rayetal}).

Finally consider the magnetic structure in the jet which is shown in
Fig.~\ref{fig:FieldAMC}.  The field structure is quite complicated as
might be expected.  Note the form of the bow-shock 
in the $B_{pol}$ component as well as its the relative absence in the
peeled off outer collar which is dominated by toroidal
fields. Examination of $\beta$ in these regions shows that the gas
remains hydrodynamically dominated with $\beta \gg 1$ 
in spite of amplification from both the shocks and
radial stretching.  Within the core the pinch modes are clearly seen
in the toroidal component with specific islands in the beam
corresponding to regions of strong pinch.  
Numerous islands of $B_{pol}$ are created by the instabilities
indicating the presence of reconnection.

\subsubsection{Multi-component Rotator:  Isothermal/Radiative Jet}

In Fig.~\ref{fig:DenRdMC} we present gray-scale maps of the density
evolution of a radiative jet driven by a {\em Multi-component}
rotator.  The radial distributions were scaled down for all variables
except $v_z$ in the jet such that $M_f \approx 9$.  The basic features
of the simulation are similar to the isothermal model however there are
some differences.  To address these we also present in
Fig.~\ref{fig:DenIsMB} a single frame from the isothermal simulation.
Note first that, once again, the width of the bow-shock is reduced in
both the radiative and isothermal case relative to adiabatic model.
This can be attributed both to cooling and the increase in fast mode
Mach number. 

The most important point to notice in this simulation is that the outer
collar still peels away from the jet core which then propagates ahead
of the rest of the flow.  This occurs even though the cooling is
strong.  As in the adiabatic model, the initiation of transverse flow
in the outer layers occurs due to the shape of the jet shock.  As the
first frame of Fig.~\ref{fig:DenRdMC} demonstrates, with cooling
included both the bow-shock and jet-shock effectively ``drape'' around
the head of the jet.  This feature occurs due to the loss of pressure
support behind the shocks.  As in the adiabatic model the peel-off
appears to be primarily driven by the redirection of the flow behind
the oblique jet-shock.  Note that, in spite of cooling, the jet-shock
in the low density collar (where $\eta < 1$) must sink back into the
body of the jet.  The bow-shock follows suit and the result is a highly
oblique section of the shock in the inner collar.  When undisturbed
beam material impinges on this shock it is either directed towards the
axis forming a strong pinch in the core or it is shunted radially
outward forcing the outer collar to peel away.  Thus in both the
adiabatic, isothermal and radiative cases the non-uniformity in the
jet-shock drives a flow pattern which enhances the 
``jet within a jet'' 
nature of the flow and the core always ends up propagating away from
the rest of the beam.

Unlike the adiabatic model where the peel-off region had a low zed
velocity, both the radiative and isothermal models show the point at
which the core and collar separate moves with a speed that is a 
large fraction of the beam speed. The origin of this effect appears to be the
lower pressures behind the jet-shock which causes less deceleration.
The propagation of the separation point may also be due to the reduced
width of the bow-shock and a smaller cocoon (both expected in
non-adiabatic models).

The magnetic field structure shown in Fig.~\ref{fig:FieldRMC} is similar
to what is seen in the adiabatic case.  Note, however, the strong pinch
which occurs at the point where the peel-off occurs.  The loss of gas
pressure support will also decrease $\beta$ implying that the toroidal
field can now exert a stronger influence.

The principle difference between the radiative and isothermal models
occurs in the core.  First note that it is difficult to see the
instabilities in the radiative model.  A detailed inspection of the
simulation frames shows they are present but they appear to diffuse
more rapidly than in the isothermal case.  The isothermal simulations
do show the same form of the modes occurring as in the adiabatic models
and with similar length scales. The difference between the isothermal
and radiative solutions is likely to due the greater thermalization
which occurs in the higher Mach number flow.

\section{Comparisons with Observations}

In recent observations of molecular outflows (\cite{dutreyetal},
\cite{guethguil}, \cite{guethetal}) show small linear structures just
ahead of the familiar bow-shaped shocks. Three examples of such
features are presented in Fig.~\ref{fig:obs}. These structures point
almost exactly away from the position of the protostellar
condensation.  These precursors of the bow-shock show a roughly conical
shape. As such they could trace an underlying jet which 
is propagating beyond the bow-shock.   The present simulations are suggestive 
offering an explanation for the observed structures. The fast 
``core''-jet propagates ahead of the collar and the surrounding molecular outflow.

The outer ``collar'' may be solely responsible for the larger bow-shock
structure or it may itself be embedded in a larger wide angle wind.
Thus molecular observations of conical precursors to the bow-shocks may
be a signature of density and magnetic stratification discussed in this
study. Therefore, the global evolution that we obtain for our jets, \eg
a core-collar structure could lead to common behavior for several YSO
jets, and also may help in understanding the relation between jets and
molecular outflows.

\section{Discussion and Conclusions}

We have carried out a series of simulations intended to address the
issue of MHD jet propagation.  Whereas previous studies have used
ad-hoc initial conditions we inject flows into our computational grid
derived from models of collimated jets driven by magneto-centrifugal
launching.  This strategy allows us to compare the propagation
characteristics of jets driven by different types of outflows. In
particular we have studied the propagation of jets driven by:
(1) a purely Keplerian rotator (a disk) exterior to a solid body
rotator (a star); (2) a Keplerian rotator with a sub-Keplerian
boundary layer both of which are exterior to a solid body rotator.
The former model we refer to as a {\em Keplerian} jet, the latter is
called a {\em Multi-component} model.  Our simulations follow the jets
out to observable scales. In the Keplerian jet simulations the jet
radius is $R_j = 1.5x10^{15}$ cm making the grid extend out to $3000 ~AU$.
For the multi-component jet $R_j$ is almost a factor of ten larger and
the the grid extends out $.1 ~pc$. The width of the multi-component
jet is interesting in that it yields a model with a very narrow,
dense core (a jet) surrounded by a wider lower density outflow.

Both models were calculated under the
assumption that the jets are launched under isothermal conditions.  We
have carried out simulations of the propagation of both {\em Keplerian}
and {\em Multi-component} jets under adiabatic, isothermal and
radiative conditions in order to determine the behavior of the
resulting flows with, and without, radiative losses.  We note again
that our adiabatic and isothermal simulations have low magneto-sonic
Mach numbers $M_{ms} = 2 ~-~ 4$.  While these values are small compared
with the values used in previous numerical studies of MHD jets $(M_{ms}
>10$, \cite{StoneHardee99}) they is quite similar to what has been
obtained in other studies MHD collimation of jets ($M_{ms} \approx
3$,\cite{cam182}).

Our simulations show significant differences in the propagation
characteristics for the two types of rotators.  In addition, features
are seen in both classes of jet which have not been seen in previous
models of either pure hydrodynamic or MHD jet propagation.  In all
cases it appears that the most important aspect of the flow behavior
seen in the simulations can be traced back to the annular
stratification of the jets.  In particular, the radial distributions of
density, velocity, and toroidal magnetic field appear to be the
principle causes of the new behavior seen in the simulations.  Both
{\em Keplerian} and {\em Multi-component} jets exhibit a core/collar
structure such that a high density core region exists
near the axis
surrounded by one or more lower density annuli (collar) extending out
to the jet boundary.  The strongest toroidal fields exist at the
boundary between the core and collar.

Since the momentum in the core is higher than that in the collar the
propagation characteristics of the jets are dominated by the core
pulling ahead of the collar.  The strong field surrounding the core
ensures that the two regions remain fairly distinct in terms of their
dynamics.  As the jets propagate we see the core acting as a {\it jet
within a jet}.  In the {\em Keplerian} case the relatively low density
contrast between core and collar keeps the two propagating at
relatively similar velocities.  The stratification of the magnetic
fields produces strong dynamical differences between core and collar.
All plasma flowing into the cocoon comes from the lower field strength
regions of the collar.  In the {\em Multi-component} case there exists
an extremely low density inner collar (which also has higher velocity
than the surrounding regions) and this leads to a complete separation
of core and collar.  The ``peel-off'' of the collar in the {\em
Multi-component} models is quite dramatic and occurs in  both the
adiabatic and isothermal simulations.

Our results have bearing on a number of issues.  The simplest
conclusion that can be drawn is that the structure imposed on a YSO jet
by the launching and collimation process can lead to fairly complex
propagation characteristics.  Thus our models build on and extend
the previous works which utilized only ``top-hat'' jets as initial
conditions. Our results also indicate that jets launched from different
classes of rotators may have different propagation characteristics.
It is likely that in real jet systems the dynamics is too complex to
make an isomorphic identification of a given class of rotators with
a set of observed jet morphologies. There is however the possibility
that as these kinds of studies mature  one might be able to distinguish
between different classes of MHD launching models via consideration
of the way the jets from these models would appear on the sky.

Finally we note that given the large parameter space of initial
conditions for both the Given Geometry Model and for the jet
propagation simulations, the work described here which focuses only on
two instances must be seen as preliminary.  It does however point to
the fact that the jets produced by magnetized rotators are likely to be
more complex in their structure and, furthermore, that this complexity
will be reflected in the observed jet morphologies.  In future studies
we will attempt to build a larger catalog of jet propagation
characteristics through a more thorough exploration of parameter space
of the Given Geometry Model.

\acknowledgements

We wish to thank Guy Delemarter, Jack Thomas, Colin Norman and Lee
Hartmann for their input and discussions leading to this paper.  This
work was supported by NSF Grant AST-0978765. DR was supported in part
by KOSEF through grant 981-0203-0011-2.

\begin{deluxetable}{l|cccccccc}
\tablecaption{\label{table1} Simulation Parameters.}
\tablewidth{0pt}
\tablehead{
\colhead{Simulation} & \colhead{$R_j ~(cm)$} & \colhead{$n ~(cm^{-3})$} &
\colhead{$\textrm{V}_{j}$ (\kms)}  & \colhead{$T_j ~(^oK)$} & \colhead{$B_\phi ~(G)$} 
}
\startdata
Adiabatic Keplerian  & $1.36\times 10^{15}$ & 120 & 120 & $3.8 \times 10^4$ & $9.1 \times 10^{-5}$  \\
Radiative Keplerian  & $1.36\times 10^{15}$ & 120 & 120 & $1.0 \times 10^4$ & $4.5 \times 10^{-5}$  \\
Adiabatic Multicomponent & $1.68\times 10^{16}$ & 120 & 190 & $3.8 \times 10^4$ & $1.2 \times 10^{-4}$  \\
Radiative Multicomponent & $1.68\times 10^{16}$ & 120 & 190 & $1.0 \times 10^4$ & $6.5 \times 10^{-5}$  
\enddata
Note: values given are maximums in jet.
\end{deluxetable}

\clearpage
\epsscale{.6}
\plotone{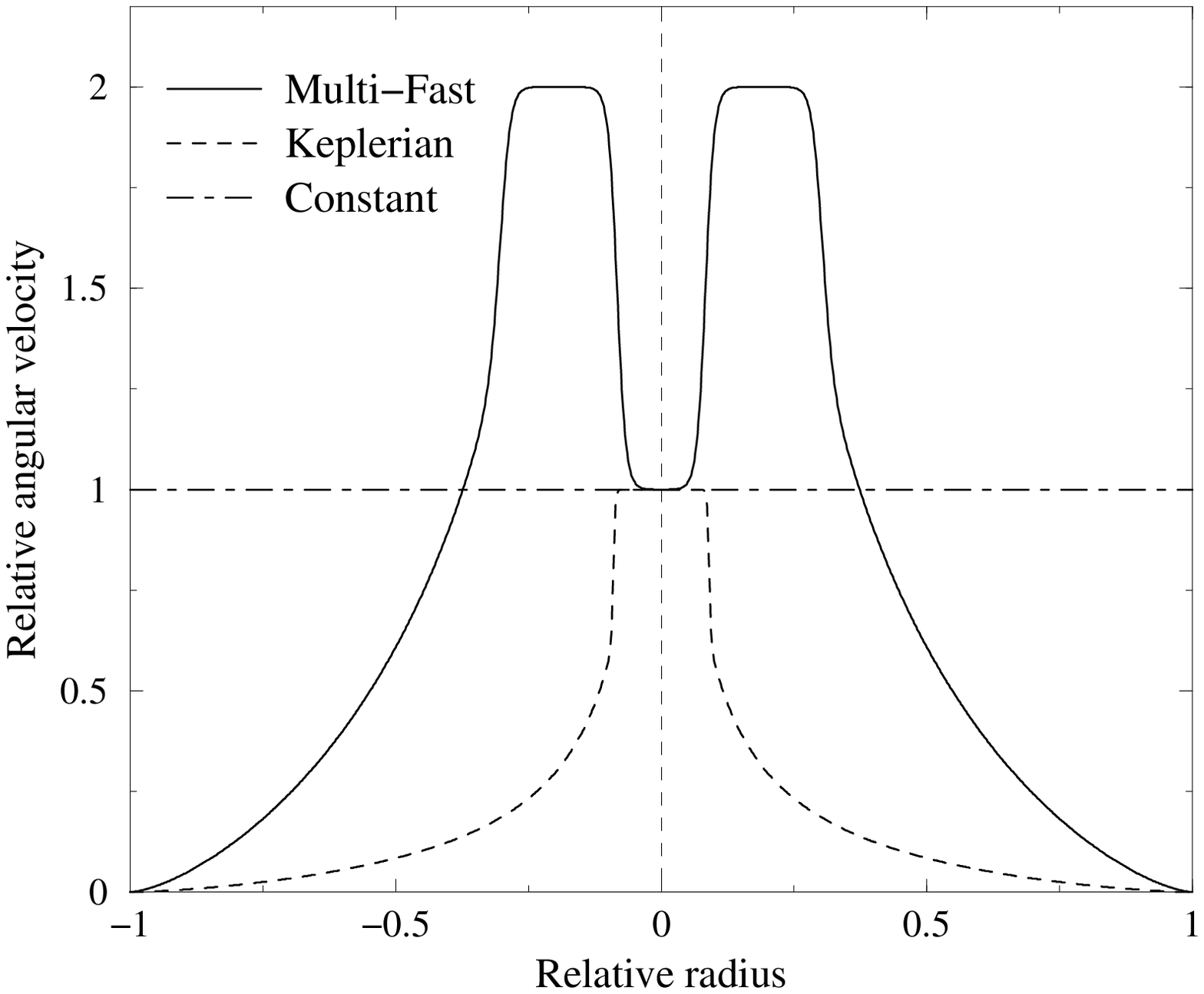}
\figcaption[]{Rotation laws for pure {\em Keplerian} (dashed), 
and {\em Multi-component} (solid) models. Axial angular velocity 
is set to unity.
\label{fig:OMEGA}}

\plotone{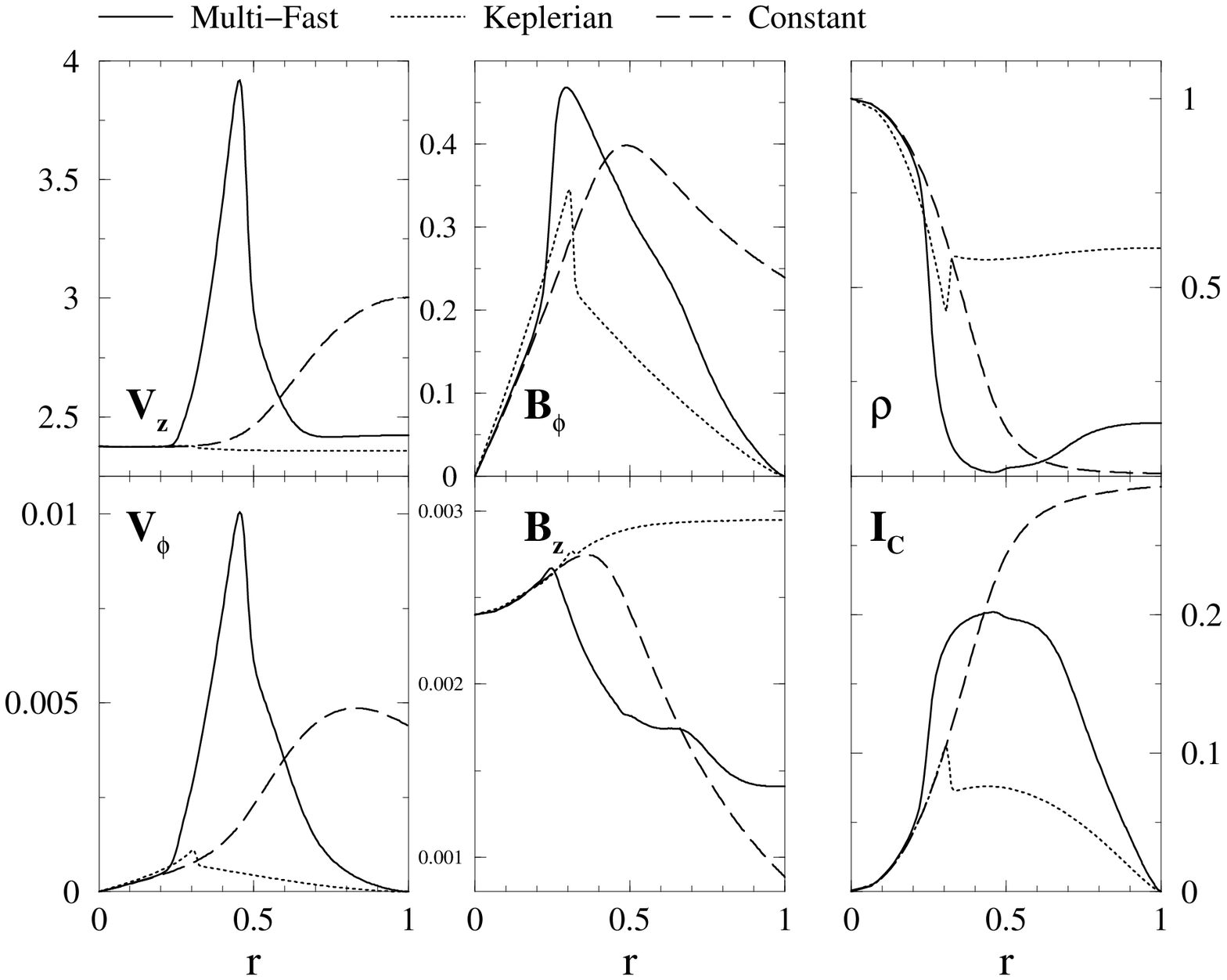}
\figcaption[]{Variations with the relative radius of the velocity $V$
and magnetic field $B$ components, of the density $\rho$ and of the
net electric current $I_C$ in the {\em cylindrically collimated regime}.
Pure {\em Keplerian} (dashed), and {\em Multi-component} (solid) rotation laws 
are considered.
\label{fig:ASYMP}}

\epsscale{1.}
\plotone{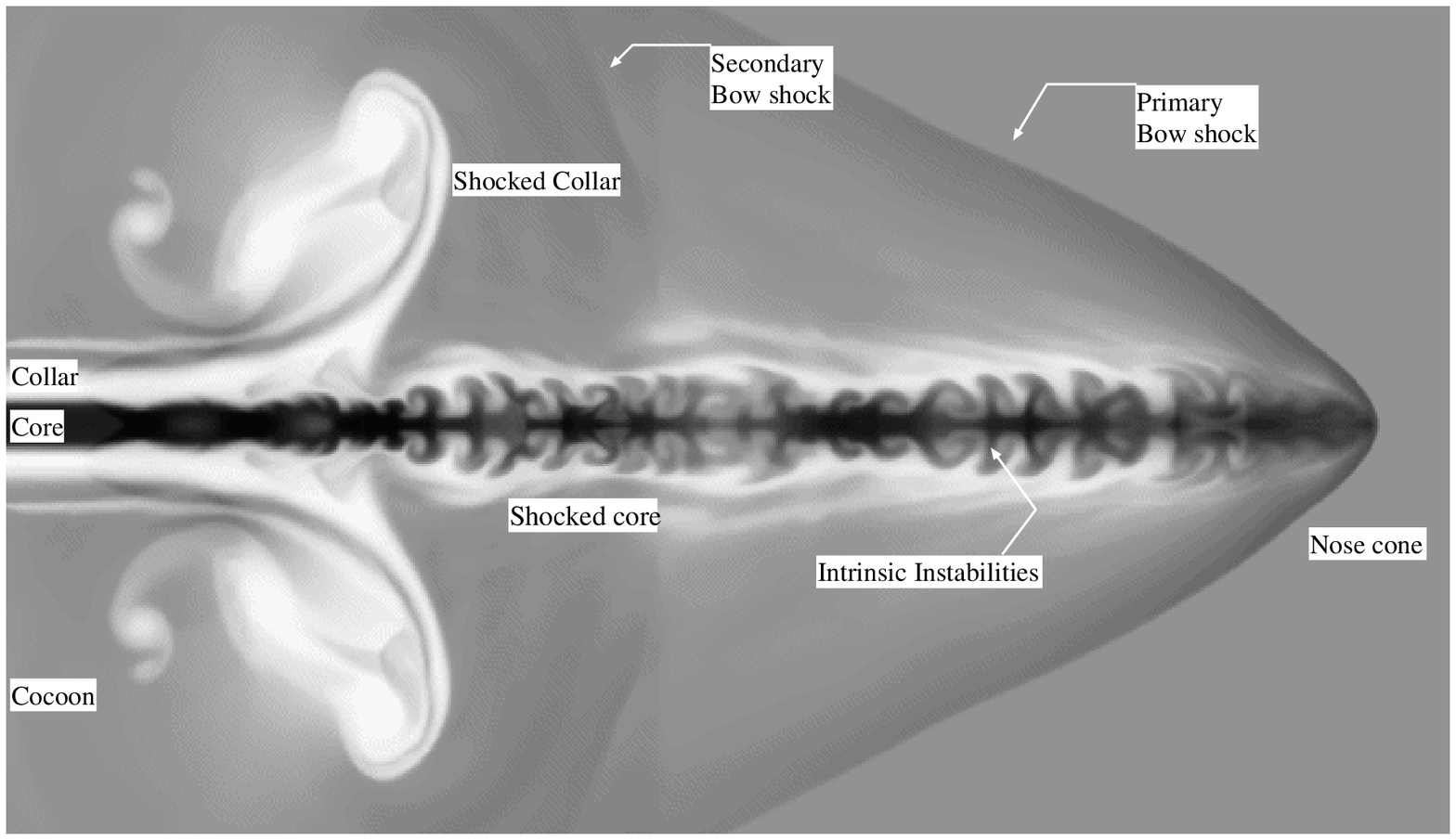}
\figcaption[]{Annotated grey-scale map of the density showing the 
basic features that are common to most of the simulations.
This corresponds to the adiabatic simulation of the {\em Multi-component}
jet at time $t=8.6$.
\label{fig:cartoon}}

\clearpage
\epsscale{.8}
\plotone{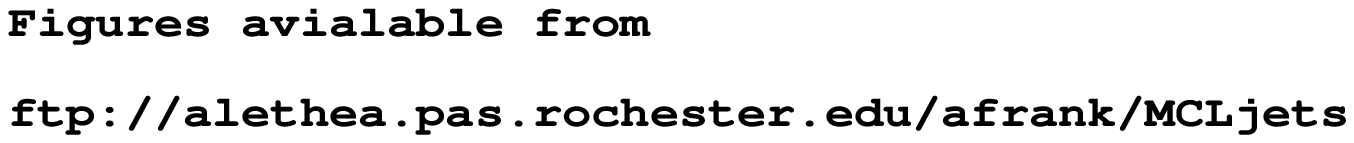}
\figcaption[]{Grey-scale maps of density for adiabatic simulations of
{\em Keplerian} jet.  Four frames from the simulation are shown.  From top
to bottom the times are $t = 5.3,10.6, 15.9, 20.2$.  
The horizontal ($Z$) and vertical
($2R$) dimensions of the simulation are $Z = 32 R_j$ 
and $R = 8 R_j$ respectively. 
\label{fig:DenAdKep}}

\clearpage
\epsscale{.9}
\plotone{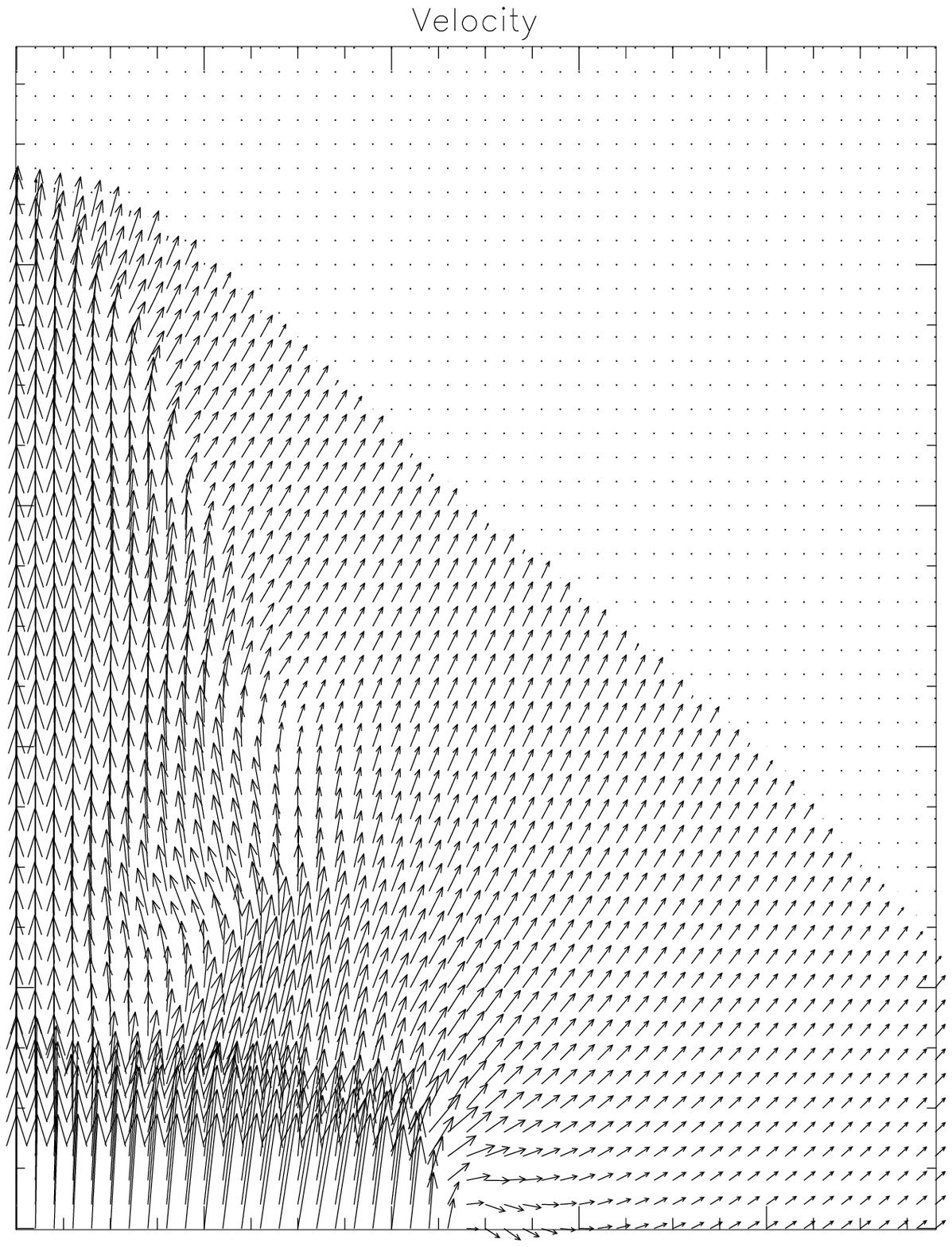}
\figcaption[]{Poloidal velocity vectors at the head of adiabatic 
Keplerian
jet.  The longest vectors correspond to a speed of $v = 99 $ km/s.  The
figure is taken at $t = 5.3$ and the size is $3.8 R_j$ on
each side. 
\label{fig:KAvhead}}

\clearpage 
\epsscale{0.9}
\plotone{afFigs.eps} 
\figcaption[]{ Magnetic field for
{\em Keplerian} Jet. Adiabatic simulation at $t = 20.2$. The top half contour
plot represents the poloidal field ($\vec{B}_p = B_r \hat{e}_r + B_z
\hat{e}_z$).  The bottom half contour plot represents the toroidal
field ($B_\phi$).  Note the increase in the magnetic pitch
$B_\phi/|B_p|$ behind shocks.
\label{fig:FieldAKep}}

\clearpage
\epsscale{.8}
\plotone{afFigs.eps}
\figcaption[]{Grey-scale maps of density for radiative simulations of
{\em Keplerian} jet.  Four frames from the simulation are shown.  From
top to bottom the times are $t = 1.6, 3.5, 5.9, 8.6$.
\label{fig:DenRdKep}}

\clearpage
\epsscale{.9}
\plotone{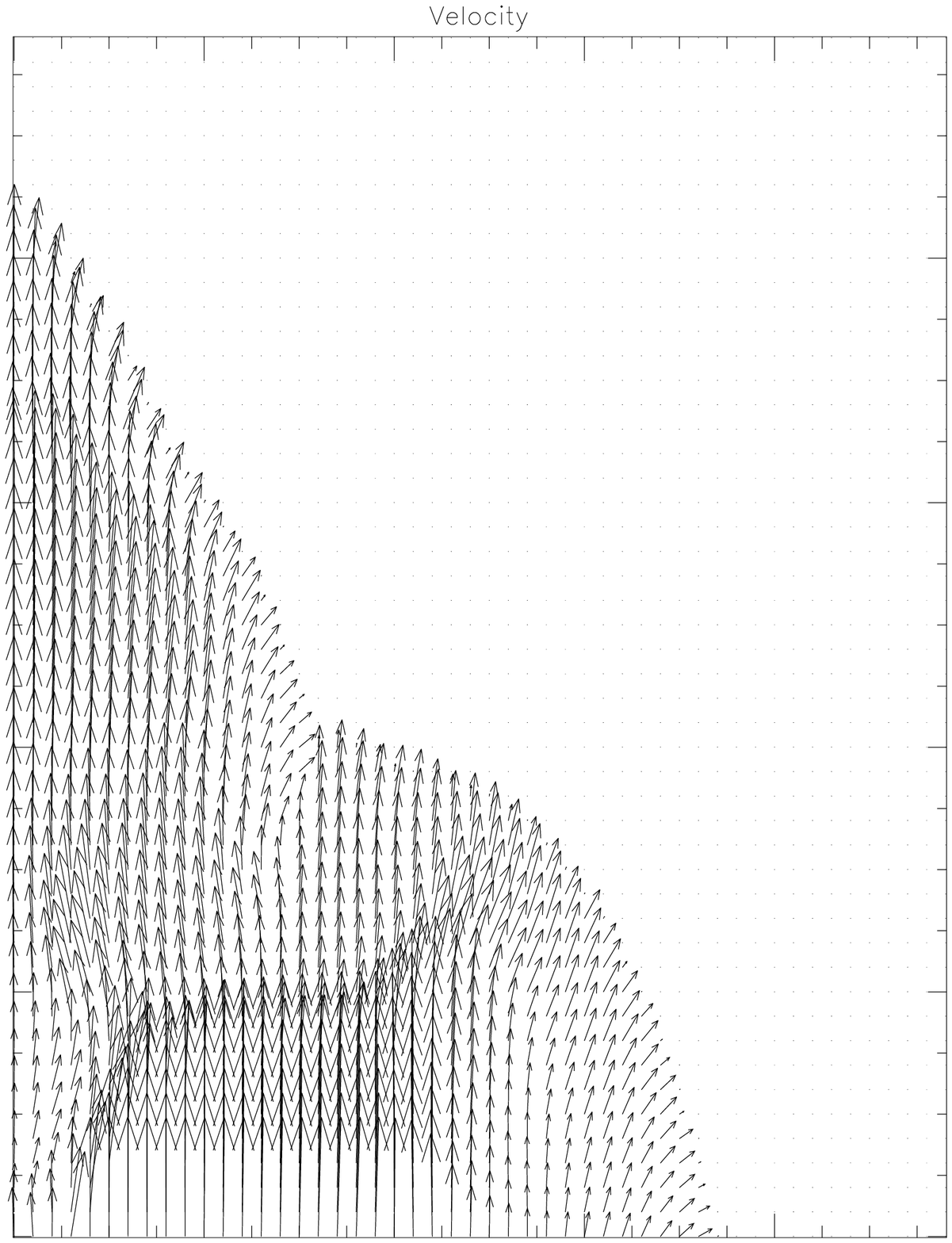}
\figcaption[]{Poloidal velocity vectors at the head of radiative
Keplerian jet.  The longest vectors correspond to a speed of $v = 120$
km/s.  Note that the largest radially inward directed flow occurs
where $\beta = \beta_{min}$. The figure is taken at $t = 5.9$ and the
physical size is $3.8 R_j$ in vertical direction.
\label{fig:KCvhead}}

\clearpage 
\epsscale{0.9}
\plotone{afFigs.eps} 
\figcaption[]{ Magnetic field for
{\em Keplerian} Jet. Radiative simulation at $t=8.6$. The top half contour
plot represents the poloidal field ($\vec{B}_p = B_r \hat{e}_r + B_z
\hat{e}_z$).  The bottom half contour plot represents the toroidal
field ($B_\phi$).  Note the increase in the magnetic pitch
$B_\phi/|B_p|$ behind shocks.
\label{fig:FieldRKep}}

\clearpage
\epsscale{.8}
\plotone{afFigs.eps}
\figcaption[]{Grey-scale maps of density for adiabatic simulations of
{\em Multi-component} jet.  
Four frames from the simulation are shown.  From top
to bottom the times are $t = 2.9, 5.7, 8.6, 11.4$ . 
 The horizontal ($Z$) and vertical
($2R$) dimensions of the simulation are $Z = 32 R_j$ and $R = 8 R_j$
respectively. 
\label{fig:DenAdMC}}

\clearpage
\epsscale{.9}
\plotone{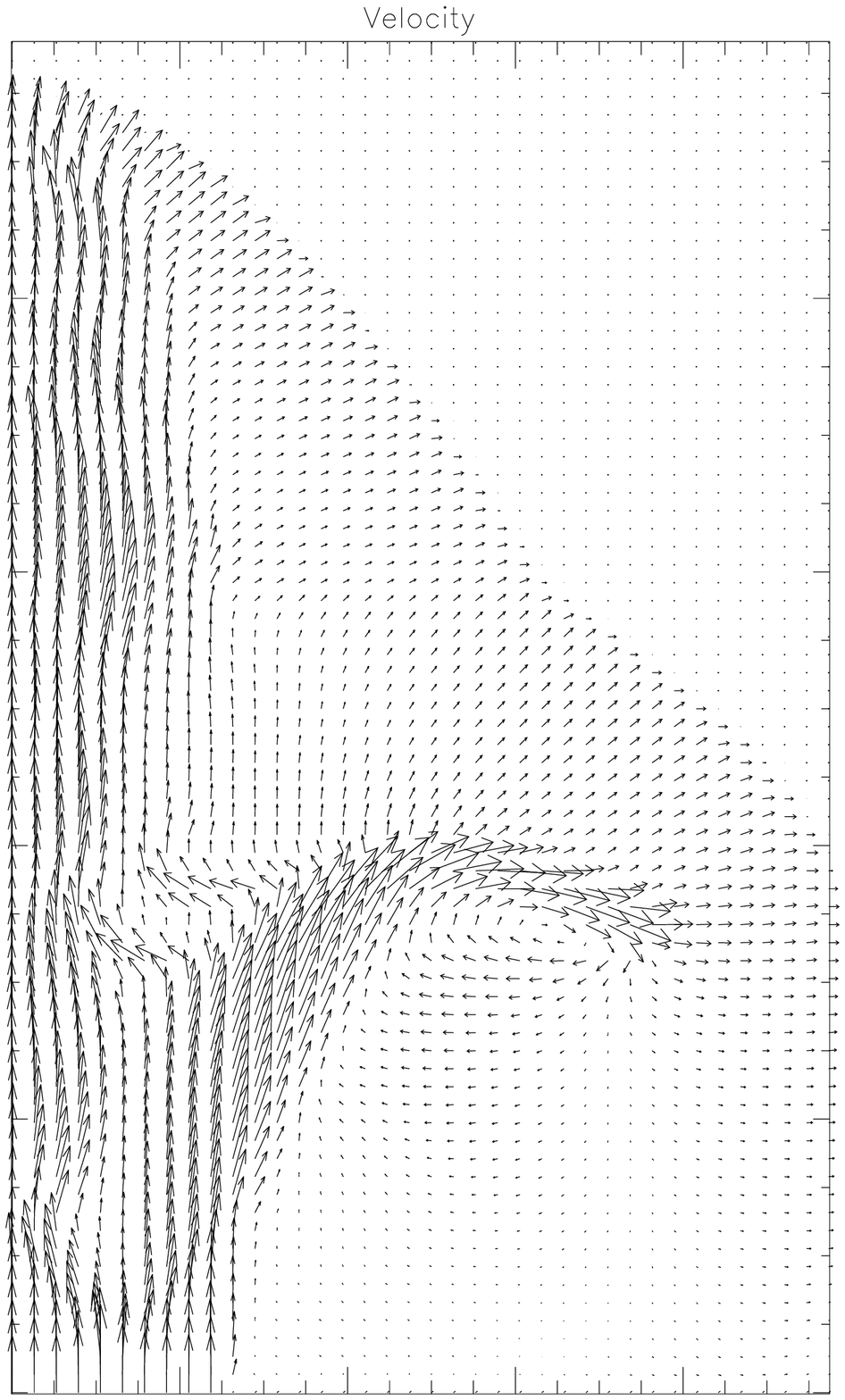}
\figcaption[]{Poloidal velocity vectors at the head of adiabatic 
{\em Multi-component} jet.  The figure is taken at $t = 2.9$ y and the
physical size is $4.5 R_j$ in vertical direction.The longest vectors
correspond to a speed of $v = 120 $ km/s.
\label{fig:MAvhead}}

\clearpage
\epsscale{.8}
\plotone{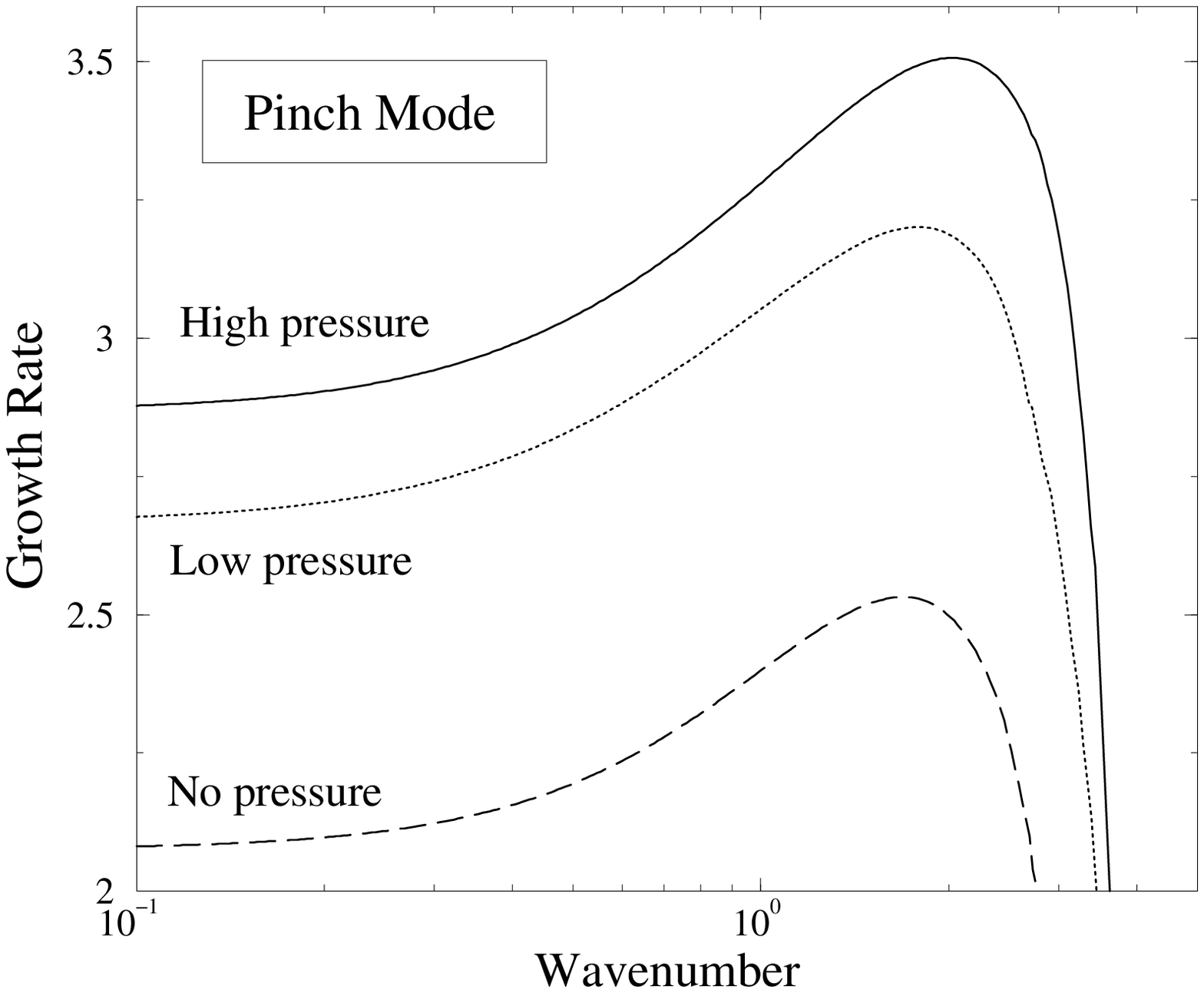}
\figcaption[]{
Stability Analysis. Dispersion relation, \ie growth rate vs.
wavenumber, for the pinch mode with different values of 
the external pressure surrounding a {\em Multi-component} jet.
\label{fig:stab}}

\clearpage
\epsscale{.9}
\plotone{afFigs.eps}
\figcaption[]{Magnetic field for {\em Multi-component} Jet. 
Adiabatic simulation.  Top half contour plot represents the poloidal field ($\vec{B}_p =
B_r \hat{e}_r + B_z \hat{e}_z$).  The bottom half contour plot
represents the toroidal field ($B_\phi$).  Note the increase in the
magnetic pitch $B_\phi/|B_p|$ behind shocks.
\label{fig:FieldAMC}}

\clearpage
\epsscale{.8}
\plotone{afFigs.eps}
\figcaption[]{Grey-scale maps of density for radiative simulations of
{\em Multi-component} jet.  Four frames from the simulation are shown.
\label{fig:DenRdMC}}

\clearpage
\epsscale{1.}
\plotone{afFigs.eps}
\figcaption[]{Magnetic field for {\em Multi-component} Jet. 
 Radiative simulation. Top half contour plot represents the poloidal field ($\vec{B}_p =
B_r \hat{e}_r + B_z \hat{e}_z$).  The bottom half contour plot
represents the toroidal field ($B_\phi$).  Note the increase in the
magnetic pitch $B_\phi/|B_p|$ behind shocks.
\label{fig:FieldRMC}}

\epsscale{.8}
\plotone{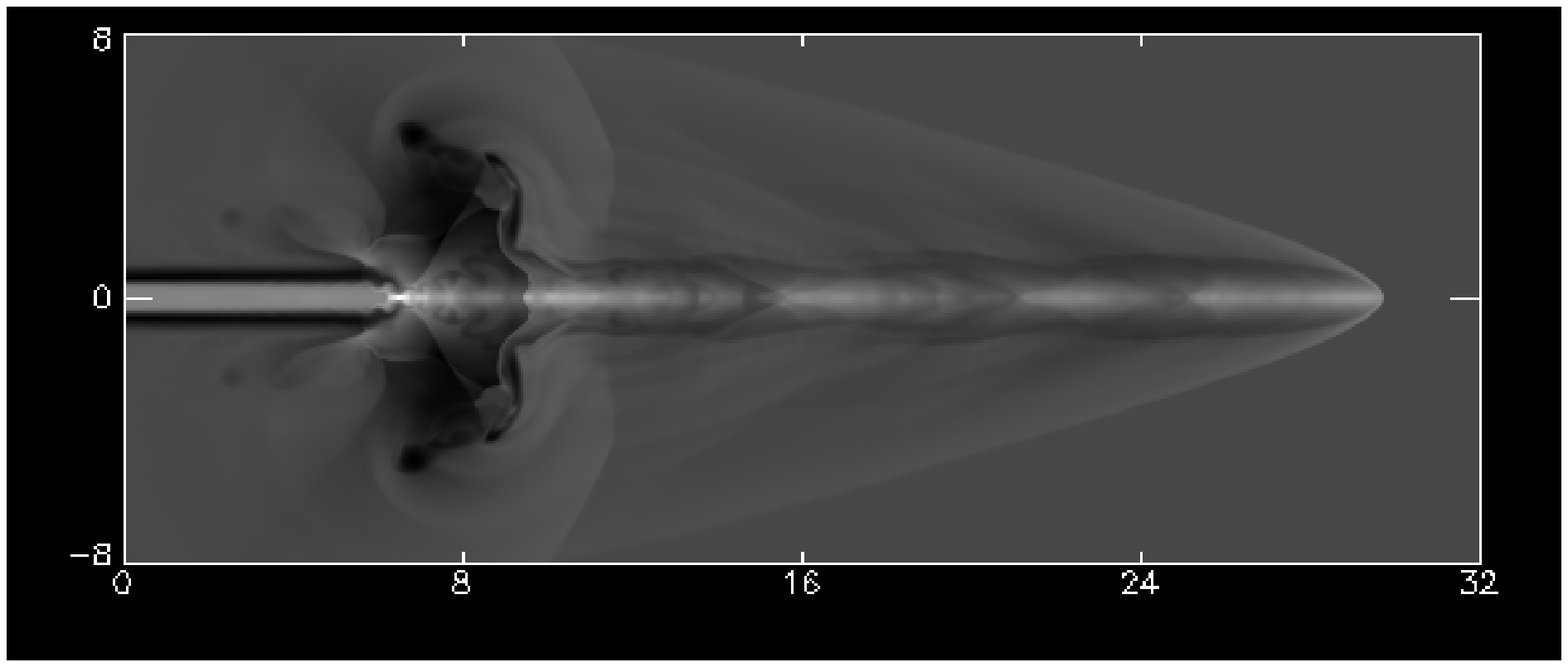}
\figcaption[]{Grey-scale map of density for isothermal simulation of
{\em Multi-component} jet.  The simulation is shown at time $t = 11.4$. 
\label{fig:DenIsMB}}

\clearpage
\epsscale{.8}
\plotone{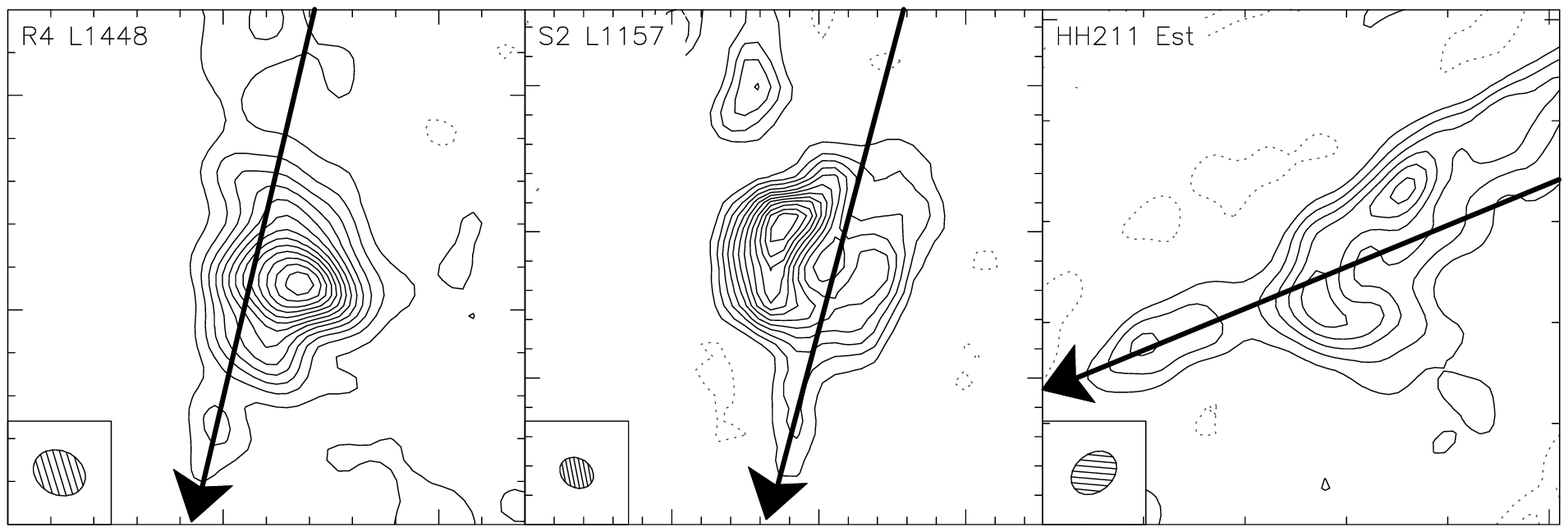}
\figcaption[]{Observations of linear precursors: 
SiO(2-1) emission of L1448 (from \cite{dutreyetal}), 
SiO(2-1) emission of L1557 (from \cite{guethetal}),
CO(2-1)  emission of HH211 (from \cite{guethguil}),
(Courtesy Gueth). The shocks exhibit an extension downstream, 
pointing away from the protostellar positions (thick lines).
Arrow head marks direction of propagation.
\label{fig:obs}}


\begin{thebibliography}{}

\bibitem[Appl, Lery \& Baty 2000]{alb}
Appl, S. Lery, T., \& Baty, H., 2000, A\&A, in press

\bibitem[Bertout \ea 1988]{bertout}
Bertout C., Basri G., Bouvier J., 1988, \apj, 330, 350

\bibitem[1988]{burke}
Burke T., Mundt R., Ray T.P., 1988, A\&A 200, 99

\bibitem[Camenzind 1997]{cam182}
Camenzind M., 1997, IAUS 182, 241

\bibitem[Cerqueira \ea 1998]{Cerqueira98} Cerqueira A. H., Gouveia Dal 
Pino E. M., \& Herant M., 1998, \apj, 489, L185

\bibitem[Cerqueira \ea 1999]{Cenquria99} Cerqueira A. H., Gouveia Dal 
Pino E. M., \& Herant M., 1999 \apj, in press

\bibitem[Dalgarno \ea 1972]{Dalgarno1972} Dalgarno A. \& McCray R. A.
1972, ARA\&A 10, 375

\bibitem[Dutrey \ea 1997]{dutreyetal} Dutrey A., Guilloteau S.,
Bachiller R., 1997, A\&A 325, 758

\bibitem[1996]{fendt}
Fendt C., Camenzind M., 1996, A\&A 313, 591

\bibitem[1997]{ferr3}
Ferreira J., 1997, A\&A 319, 340

\bibitem[Frank \ea 1998]{Frankea98} Frank A., Ryu D., Jones T. W. \&
Noriega-Crespo A. 1998, \apj, 494, L79

\bibitem[Frank \ea 1999]{Frankea99} Frank A., Gardiner, T.,
Delamarter, G., \& Betti R., 1998, \apj, in press

\bibitem[Gardiner \ea 1999] {Gardinerea99}
Gardiner T., Frank A., Ryu D., Jones T., 1999, \apj, submitted

\bibitem[Gardiner \& Frank 1999] {GardFrank99}
Gardiner T., \& Frank A.,  1999, \apj, in preparation

\bibitem[Goodson \ea 1997]{Goodsonea97}
Goodson A, Winglee R., Boehm K., 1997, \apj, 489, 199

\bibitem[Gueth \ea 1998]{guethetal}
Gueth F., Guilloteau S.,  1998, A\&A 333, 287 

\bibitem[Gueth \& Guilloteau 1997]{guethguil}
Gueth F., Guilloteau S.,  1997, A\&A 325, 758

\bibitem[Hartmann 1998] {Hartmann98}
Hartmann L., 1998, in Accretion Processes in Star Formation,
(Cambridge U. Press)

\bibitem[Harten 1983]{Harten1983} Harten A., 1983, J.Comp.Phys. 49, 357

\bibitem[Heyvaerts \& Norman 1989]{HevNor89}
Heyvaerts J., Norman C.A.,  1989, \apj, 347, 1055

\bibitem[Hollenbach \& McKee 1979]{HollMcKee79}
Hollenbach D., \& McKee R., 1979, \apj, 41, 555

\bibitem[Kudoh \ea 1998]{Kudohea98}
Kudoh T., Matsumoto R., Shibata K.,  1998, \apj, 508, 186

\bibitem[Leahy 1991] {Leahy91}
Leahy J.P., 1991, in ``Beams and Jets in Astrophysics'', ed. P. Hughes, (Cambridge U. Press)

\bibitem[Lery 1996]{lery3}
Lery T., 1996, Ph.D. Thesis, ULP Strasbourg, France

\bibitem[1999b]{LHF}
Lery T.,  Henriksen R.N., Fiege J.D., 1999b, A\&A, in press

\bibitem[Lery \ea 1998] {Leryea98}
Lery T., Heyvaerts J., Appl S., Norman C.A., 1998, A\&A 337, 603

\bibitem[Lery \ea 1999] {Leryea99}
Lery T., Heyvaerts J., Appl S., Norman C.A., 1999, A\&A, 347, 1055

\bibitem[Lery \& Frank 1999] {LeryFrank99}
Lery T. \& Frank, A., 1999, \apj, submitted

\bibitem[Leveque 1998] {Leveque98}
Leveque, Randall J. 1998, in Computational Methods for Astrophysical
Fluid Flow, (Springer, Berlin, New York), p.1

\bibitem[Lind \ea 1989] {Lindea89}
Lind K., Payne D., Meier D., \& Blandford R., 1989, \apj, 344, 89

\bibitem[1997]{ostriker}
Ostriker E.C., 1997, \apj, 486, 291

\bibitem[Ouyed \& Pudritz 1997]{OP97a} Ouyed R. \& Pudritz R. E. 1997a, 
\apj, 482, 712

\bibitem[Priest 1986] {Priest86}
Priest E. 1986, Solar Magnetohydrodynamics 

\bibitem[Pudritz 1991]{Pudritz91}
Pudritz R.E. 1991, in ``The Physics of Star Formation and Early Stellar
Evolution'', eds. C.J. Lada and N.D. Kylafis, NATO ASI Series (Kluwer), 365

\bibitem[Ray \ea 1997]{Ray1997} Ray T. P., Muxlow T. W. B., Axon D. J., 
Brown A., Corcoran D., Dyson J. \& Mundt R. 1997, Nature, 385, 415

\bibitem[Ramm 1990] {Ramm90} Ramm H., 1990, 
``Fluid Dynamics for the Study of Transonic Flow''
(Oxford U. Press, New York) 

\bibitem[1996]{rayetal}
Ray T.P., Mundt R., Dyson J.E., Falle S.A.E.G.,
Raga A.C., 1996, \apj, 468, L103

\bibitem[Reipurth 1997]{Reipurth97}
Reipurth B., 1997, in Herbig-Haro Flows and the Birth of
Low Mass Stars, in IAU Symposium no. 182, eds B. Reipurth \& C Bertout
(Kluwer, Dordrecht)

\bibitem[Romanova \ea 1998]{Romanova98} Romanova M. M., Ustyugova, G. V., 
Koldoba A.V., Chechetkin V.M., \& Lovelace R.V.E. 1998, \apj, 
500, 703

\bibitem[Ryu \& Jones 1995]{Ryu1995-1D} Ryu D.\& Jones T. W. 1995, \apj,
442, 228

\bibitem[Ryu \ea{} 1995a]{Ryu1995-2Dxy} Ryu D., Jones T. W., Frank A. 1995,
\apj, 452, 785

\bibitem[Ryu \ea{} 1995b]{Ryu1995-2Dcyl} Ryu D., Yun H. S., Choe S. 1995,
J. Korean Ast. Soc. 28, 243

\bibitem[Ryu \ea{} 1998]{DivBmeth} Ryu D., Miniati F., Jones T. W., \&
Frank A. 1998, \apj, 509, 244

\bibitem[1994]{sautytsing}
Sauty C., Tsinganos K., 1994, A\&A 287, 893

\bibitem[Shu {\it et al.} 1994] {Shuea94}   
Shu F., Najita J., Ostriker E., Wilkin F., Ruden S., \& Lizano S.,
1994, \apj, 429, 781

\bibitem[1995]{shu95}
Shu F.H., Najita J., Ostriker E., Shang H., 1995, \apj, 455, L155

\bibitem[Soker \& Livio 1994]{SokLiv94}
Soker N., \& Livio M., 1994, AJ 421, 219

\bibitem[Stone \& Hardee 1999] {StoneHardee99}  
Stone J, \& Hardee P., 1999, preprint

\bibitem[Strang 1968]{Strang} Strang G., 1968, SIAM J. Numer. Anal.
5, 506

\bibitem[Todo \ea 1992] {Todo92}
Todo Y., Uchida Y., Sato T., \& Rosner R., 1992, PASJ 44,245

\bibitem[1997]{trussoni}
Trussoni E., Tsinganos K., Sauty C., 1997, A\&A, 325,1099

\bibitem[Vlahakis \& Tsinganos 1998]{VT}
Vlahakis, N.\& Tsinganos,K, 1998, MNRAS, 298, 777

\end{thebibliography}
\end{document}